# Forging the Industrial Metaverse: Where Industry 5.0, Augmented/Mixed Reality, IIoT, Opportunistic Edge Computing and Digital Twins Meet

TIAGO M. FERNÁNDEZ-CARAMÉS[1,2] (Senior Member, IEEE), PAULA FRAGA-LAMAS[1,2] (Senior Member, IEEE)
[1] Department of Computer Engineering, Faculty of Computer Science, Campus de Elviña s/n, Universidade da Coruña, 15071, A Coruña, Spain
[2] Centro de Investigación CITIC, Universidade da Coruña, 15071 A Coruña, Spain

Corresponding authors: Tiago M. Fernández-Caramés and Paula Fraga-Lamas (e-mail: tiago.fernandez@udc.es, paula.fraga@udc.es).

This work has been funded by grant PID2020-118857RA-100 (ORBALLO) funded by MCIN/AEI/10.13039/501100011033.

**ABSTRACT** The Metaverse, which in the last years has already become a buzzword, is a concept that proposes to immerse users into real-time rendered 3D content virtual worlds delivered through Extended Reality (XR) devices like Augmented and Mixed Reality (AR/MR) smart glasses and Virtual Reality (VR) headsets. When the Metaverse concept is applied to industrial environments, it is called Industrial Metaverse, a hybrid world where industrial operators work by using some of the latest technologies. Currently, such technologies are related to the ones fostered by Industry 4.0, which is evolving towards Industry 5.0, a paradigm that enhances Industry 4.0 by creating a more sustainable and resilient world of industrial human-centric applications. The Industrial Metaverse can benefit from the concepts fostered by Industry 5.0, since it implies making use of dynamic and up-to-date content, as well as fast human-to-machine interactions. To enable such enhancements, this article proposes the concept of Meta-Operator, which is essentially an industrial worker that follows the principles of Industry 5.0 and interacts with Industrial Metaverse applications and with his/her surroundings through advanced XR devices. In order to build the foundations of future Meta-Operators, this article provides a thorough description of the main technologies that support such a concept: the main components of the Industrial Metaverse, the latest XR technologies and accessories and the use of Opportunistic Edge Computing (OEC) communications (to detect and interact with the surrounding Internet of Things (IoT) and Industrial IoT (IIoT) devices). Moreover, this paper analyzes how to create the next generation of Industrial Metaverse applications based on the Industry 5.0 concepts, including the most relevant standardization initiatives, the integration of AR/MR devices with IoT/IIoT solutions, the development of advanced communications and software architectures and the creation of shared experiences and opportunistic collaborative protocols. Finally, this article provides an extensive list of potential Industry 5.0 applications for the Industrial Metaverse and analyzes thoroughly the main challenges and research lines. Thus, this article provides a holistic view and useful guidelines for the future developers and researchers that will create the next generation of applications for the Industrial Metaverse.

## I. INTRODUCTION

Extended Reality (XR) technologies like Augmented Reality (AR) and Mixed Reality (MR) have evolved significantly since the 1960s, when the first pioneering solutions were built [1], [2]. In fact, there was not a significant progress in AR and MR until the late 90s, when academic [3] and industrial works [4], [5] pushed the technologies again. Thus, such a progress led in the last years to the use of AR and MR in multiple industrial manufacturing processes [6], [7], [8] and specific industries (e.g., automotive industry [9],[10], [11]).

AR and MR are two of the essential technologies of the Industry 4.0 paradigm [12], [13], since they allow for improving factory performance [14], [15]. A step farther than Industry 4.0 is Industry 5.0, which has been recently characterized by the European Commission [16]. Such a new paradigm was conceived to address some of the concerns raised by Industry 4.0 in relation to social fairness and sustainability. Specifically, Industry 5.0 goals are aimed at providing sustainable manufacturing and operator well-being [17] through three main core values: sustainability, resilience and human centricity. This article essentially focuses on the latter aspect (the development of AR/MR human-centric applications), but, as it is depicted in Figure 1, the future Industry 5.0 operator (called in this paper Meta-Operator), will also have to consider resilience and sustainability, while being able to make use of Industry 4.0 technologies.

A complementary concept to Industry 5.0 is the Industrial Metaverse, which can be defined as a network of real-time rendered 3D virtual worlds related to industrial applications [18]. Such a concept has in common with the Industry 5.0 paradigm the fact that one of its important objectives is to develop human-centric AR/MR industrial applications. Although the Industrial Metaverse is still in its infancy, it is expected to have a significant economic impact, with a market size that several studies value between $22 and $540 billion [19].

Although there is an increasing number of AR/MR solutions, very few of them have been devised with the Industry 5.0 paradigm and the Industrial Metaverse in mind and thus provide shared experiences or real-world manipulation of Internet of Things (IoT) or Industrial Internet of Things (IIoT) objects. Such shared experiences add the possibility to immerse multiple users in the same AR/MR scenario in a way that they can interact simultaneously and with the same virtual elements. Moreover, advanced shared experiences integrate IoT/IIoT systems with AR/MR, thus allowing virtual and real elements to remain synchronized and to react to the changes that happen in both worlds. As a consequence, response latency is a key factor in a human-centric AR/MR shared experience, since it impacts user experience (UX) by desynchronizing the visualization and reactions of virtual elements. In addition, the devices that are part of an AR/MR shared experience must share their physical location in the real world, so the use of an external server in a remote cloud is needed, but it involves a communications delay that is longer than the one required when communications are performed in a local network. Furthermore, when different groups of AR/MR users are in different physical locations while using shared virtual assets, it is necessary to transmit the shared information to the different locations with as little delay as possible. In this scenario, the use of an opportunistic edge computing-based architecture is appropriate, since it limits latency by restricting packet exchanges with the cloud, while providing fast device discovery mechanisms[20].

Considering the previously mentioned problems, this article proposes the concept of Meta-Operator, an Industry 5.0 industrial worker whose knowledge and abilities are augmented through the use of advanced AR/MR devices and that is able to interact with the entities that participate in Industrial Metaverses.

Previous similar reviews did not explicitly consider the concept of Meta-Operator and were focused in other specific topics, like the commercial Metaverse (i.e., the Metaverse for the general public) [21], [22], [23], Web 3.0 [24] or a specific industry or technology [25], [26]. Moreover, while some works analyze the impact of Industry 4.0 and the Industrial Metaverse [27], [28], only a few authors have considered Industry 5.0, but they mainly focused on the overall picture, without delving into the specifics of the involved technologies [29], [30], [31]. In contrast to the previously mentioned papers, this article focuses on providing a holistic view on three concepts (Industrial Metaverse, Meta-Operators, Industry 5.0) that together with some enabling technologies (i.e., Augmented/Mixed Reality, IIoT, Opportunistic Edge Computing and Digital Twins) will pave the way for the creation of the next generation of smart factories. It provides the following main contributions, which, as of writing, have not been found together in the literature:

- This article introduces the concept of Industry 5.0 Meta-Operator, an Industry 5.0 operator whose knowledge and abilities are enhanced through the use of advanced AR/MR devices.
- It provides a thorough comparison on the latest AR/MR smart glasses that can be used to provide XR services to Meta-Operators.
- It identifies and discusses the main aspects of AR/MR for enabling the interaction with IoT and IIoT devices, including the use of Opportunistic Edge Computing (OEC) communications.
- It provides an extensive list of relevant development, efficiency and legal challenges that future developers will have to face in the next years.
- It identifies the most promising research lines for creating the Industrial Metaverse.

The rest of this article is structured as follows. Section II reviews the background on Industry 5.0, Industrial Metaverse, XR technologies behind the metaverse, the development of opportunistic edge computing communications, and main standardization initiatives. Section III defines the Meta-Operator concept, it analyzes the most relevant AR/MR devices, it studies useful accessories for industrial meta-verse applications, it identifies industrial metaverse software platforms, it reviews previous AR/MR developments that enable the interaction with IoT/IIoT devices, and describes the main aspects of industrial digital twins. Section IV presents the design of AR/MR applications for the Industry 5.0 metaverse, including its communications and software architecture, shared industrial metaverse experiences and opportunistic collaborative protocols. Section V describes relevant Industry 5.0 applications for the industrial metaverse. Section VI outlines the main challenges, while Section VII summarizes the main research lines for the creation of the future Industrial Metaverse. Finally, Section VIII is devoted to conclusions.

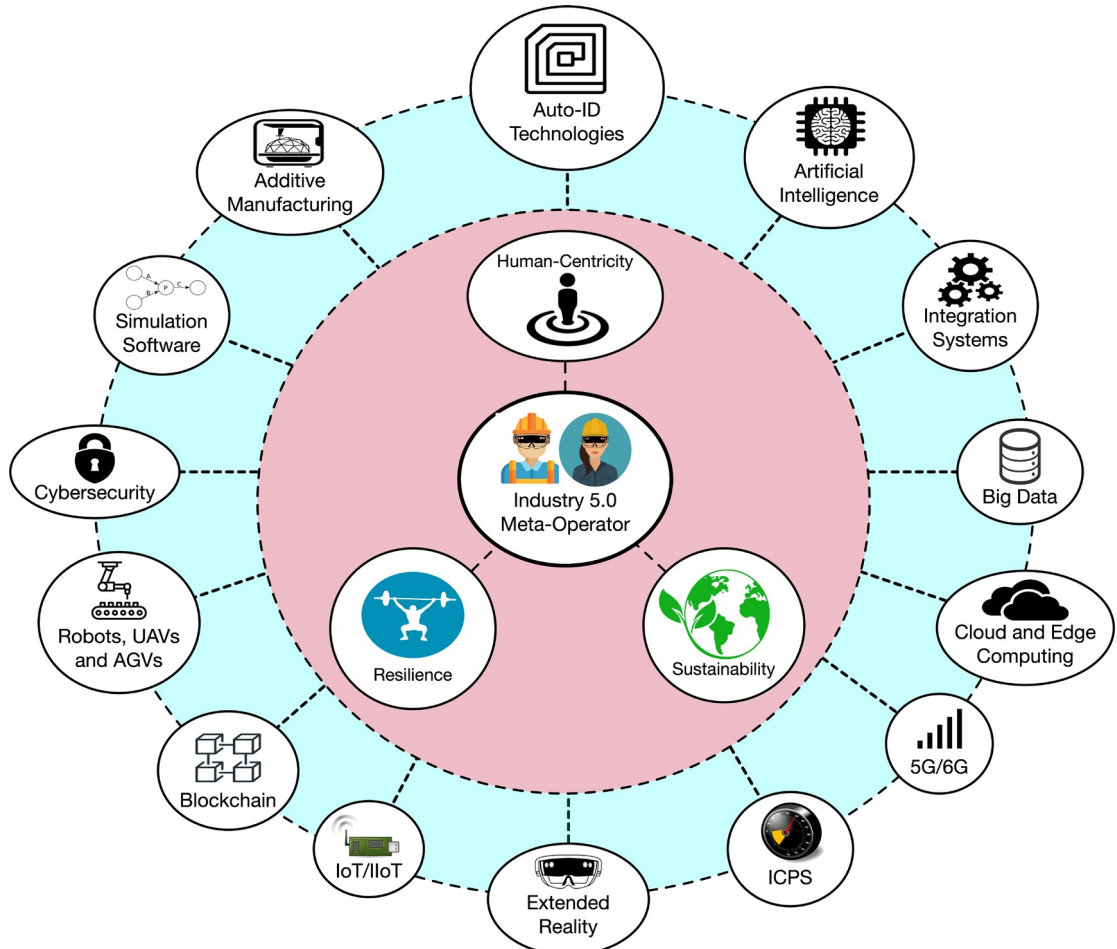

FIGURE 1: Main areas related to the Industry 5.0 Meta-Operator concept.

## II. BACKGROUND
### A. ON INDUSTRY 5.0
Industry 5.0 is a paradigm currently fostered by the European Commission that seeks to go beyond Industry 4.0 to achieve jointly economic growth, industrial progress and societal goals [16]. Therefore, Industry 5.0 aims for long-term prosperity by increasing productivity without displacing human employees from industrial businesses.

It is important to note that Industry 5.0 is neither a chronological continuation or a replacement of the Industry 4.0 paradigm [16]. Instead, Industry 5.0 may be seen as a fusion of contemporary European industrial and societal developments that complements Industry 4.0 core objectives. Since its beginnings in 2011 [32], Industry 4.0 has been mostly focused on industrial digitalization, production flexibility and efficiency optimization, rather than on societal issues like social fairness or environmental impact. As a result, Industry 5.0 refocuses Industry 4.0 principles by reorienting industrial research and innovation towards a human-centered and environmentally conscious future.

To achieve the previously mentioned goals, the European Commission identified six fields that are supposed to be essential for the future technological progress of industry [33]: individualized human-machine interaction; bio-inspired technologies and smart materials; digital twins and simulation; data transmission, storage and analysis technologies; Artificial Intelligence (AI); and technologies for energy efficiency, renewable energy, storage and autonomy. XR technologies will contribute to the first field (individualized human-machine interaction) and go farther by providing the visual interfaces required by Industrial Metaverse.

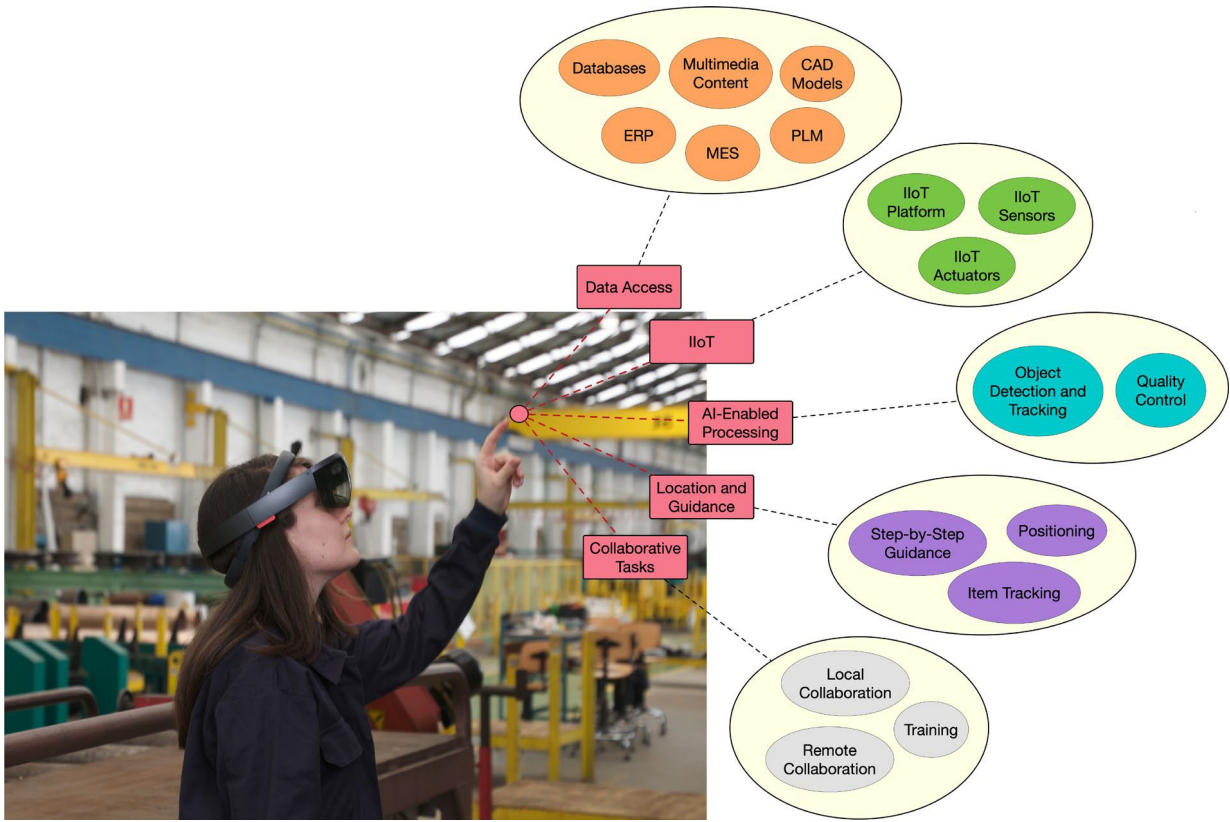

FIGURE 2: Main functionality that can be accessed by a Meta-Operator.

### B. ON THE INDUSTRIAL METAVERSE

The term 'Metaverse' originated from Neal Stephenson's 1992 novel Snow Crash [34]. Stephenson's literary contribution depicted a persistent virtual world permeating human existence, blurring the lines between labor, leisure, art and commerce. Specifically, the word 'Metaverse' derives from the Greek prefix 'meta', which means 'beyond', and the stem 'verse', which comes from 'universe'. Therefore, in English, 'metaverse' suggests transcending or going beyond our universe.

Despite this captivating concept, the Metaverse lacks a unified definition [35], [36], allowing industry leaders to shape it according to their worldviews and corporate capabilities [37], [38]. Despite the lack of consensus, the sheer number of companies recognizing potential value in the Metaverse underscores the magnitude and diversity of the opportunity [19], [39]. Within the discourse, executives frequently adopt the buzzword without a comprehensive understanding, reflecting the evolving nature of the term [37].

The debate on the Metaverse definition is also extended to the technologies that are part of its core. For instance, authors have previously discussed about whether AR is integral to the Metaverse or distinct from it [40]. Moreover, some authors consider the Metaverse as a decentralized version of the current Internet, emphasizing user control over underlying systems, data and virtual goods [41].

The ambiguous boundary between a 'metaverse' and a participatory XR environment further complicates the definition. For instance, everyday items, such as IoT-based home automation systems that enable XR interaction [42] exhibit Metaverse-like qualities, but they are usually considered Cyber-Physical Systems (CPSs) [43] rather than pure Metaverse developments. As a consequence, the term 'Metaverse' emerges as a fixture in the ever-evolving technological landscape.

The transformative power of the Industrial Metaverse lies not only in its ability to facilitate remote collaboration but also in its capacity to enhance the efficiency and effectiveness of industrial processes. From ubiquitous computing to advanced tracking technologies, the Metaverse offers a multifaceted approach to industrial applications, promising a future where technology seamlessly integrates with the workforce, generating unprecedented value and insights.

In the case of the Industrial Metaverse, it encapsulates a vision of the future where professional and commercial industrial aspects seamlessly intertwine digitally and physically. Thus, the term can be related to an interactive virtual version of the Internet, transcending traditional physical boundaries. In fact, at its core, the Metaverse represents a massively scaled and interoperable network of real-time rendered 3D virtual worlds [44]. Users experience it synchronously and persistently, fostering an individual sense of presence with continuity of data. This encompasses identity, history, entitlements, objects, communications and payments. The term extends beyond a singular entity or to a specific industry: there are numerous Metaverses catering to diverse interests such as sports, movies, art and commerce [45].

Nowadays, it seems that multiple metaverses will coexist [46], so a network of them will derive in 'metagalaxies', which are collections of virtual worlds connected under a single authority, reminiscent of the intricate structure of the Internet and the geopolitical organization of the physical world. As an example, Figure 3 shows how a remote Meta-Operator, after setting a common professional profile, would access an array of external metaverses.

It must also be noted that some authors have already suggested the term 'multiverse' instead of metagalaxy [35] and point out that many metaverses will be created in the next decade, while others consider that the Metaverse concept, as defined by companies in the last years, is actually a technology bubble that cannot be achieved in part due to being a set of loosely connected activities [47].

It is important to note that several platforms, which can be considered as 'Proto-Metaverse' platforms, such as Second Life [48], Fortnite [49], Minecraft [50] or Roblox [51], offer a glimpse into the integration of Metaverse features within collaborative applications. Specifically, in such environments, users not only engage in gameplay but also work collaboratively, attend virtual events, and participate in real-world economic transactions, exchanging money for digital goods and services within virtual marketplaces.

However, despite the immersive experiences provided by the mentioned platforms, they have largely operated as isolated universes. Therefore, no seamless transitions occur between these diverse virtual realities, so users do not maintain a consistent virtual identity, represented by an avatar, across different metaverses. Furthermore, the digital assets accumulated in one metaverse are not transferred automatically to other metaverses.

Table 1 compares the most relevant characteristics of traditional networks (i.e., current Internet and traditional industrial networks) and the new metaverses: the Commercial Metaverse for the general public and the Industrial Metaverse for the next generation of Industry 5.0 companies. As it can be observed, the Industrial Metaverse has in common certain aspects with traditional industrial networks (e.g., failure severity, required reliability, estimated latency or the need for determinism) since future Industrial Metaverses will be built on top of IoT/IIoT devices that will run on industrial networks. However, there is a clear difference between both regarding the type of content commonly exchanged (i.e., sensing/control data vs 3D content) and on the fact that Meta-Operators can work from remote locations, so they do not necessarily need to be on-site, in industrial areas where there is dust, noise or communications interference (although the underlying IoT/IIoT networks operate in such conditions).

When comparing the Commercial and the Industrial Metaverse, it can be observed that they diverge in the previously mentioned critical aspects, which are related to the fact that industrial developments need to be more robust and reliable than commercial applications. Moreover, ideally, latency should be lower when integrating data from IoT/IIoT devices, because certain sensors and actuators work at a higher update rates than most of the data sources of Commercial Metaverse solutions.

Finally, it must be emphasized that human-centricity is the key for joining Industry 5.0 and the Industrial Metaverse [52], [29]:

- The success of the Industrial Metaverse (and of the Commercial Metaverse) is conditioned by the level of provided immersiveness, which depends on adapting to the human senses (essentially to sight and hearing) and feelings (e.g., touch through haptic devices).
- In the Industrial Metaverse user experience is essential for its success: if the underlying protocols and technologies do not adapt to UX requirements (i.e., if they do not become human centric), the whole system will be a failure.
- Accessibility and inclusion are fundamental for the Industrial Metaverse, needing to provide adequate interfaces and comfortable hardware devices to gain user acceptance.
- The Industrial Metaverse should be open to anyone, avoiding past company wars, which usually ended up in vendor-locking scenarios, and, as a consequence, in limiting the growth of the developed solutions.

### C. XR TECHNOLOGIES BEHIND THE METAVERSE

The proponents of the concept of Metaverse envision it as a future 3D overlay on the real world, where individuals engage in commerce, gaming and collaborative virtual environments seamlessly [53]. Despite corporate investments (e.g., Facebook's multi-billion investment during the last years), the realization of a 'true' Metaverse remains elusive, with a myriad of companies laying the foundation for its eventual emergence [39].

Technological strides have been made towards creating immersive virtual realities, with companies like Facebook exploring Virtual Reality (VR) through advanced eyeglasses and high-quality visuals [54]. AR and MR smart glasses have also emerged, providing users with additional information overlaid on their physical surroundings [13], which anticipates a range of entry points, from smartphones to personal computers or televisions. For instance, Microsoft Azure cloud [55] and MR headsets [56] contribute to this vision, blurring the lines between virtual, mixed and augmented reality. As a consequence, virtual worlds, in this context, can range from immersive 3D environments to purely text-based augmented scenarios.

Distinguishing among the different XR technologies is crucial for understanding the Industrial Metaverse landscape. For such a purpose, Figure 4 depicts the most relevant XR technologies, whose name and definition has varied in the literature during the last decades [57]. For the sake of clarity, the following definitions will be used in this article:

- *r* Augmented Reality: it situates users within their physical environment, enhancing it with additional information or by removing part of it. Such a definition implies that virtual content can be added to reality or that such a reality can be modified either by changing it or by removing certain objects. Thus, in this article the definition of AR includes the following technologies, which are types of which is known as Mediated Reality [58]:
  – Assisted Reality (aR). It superimposes virtual content on reality with the objective of providing ad-

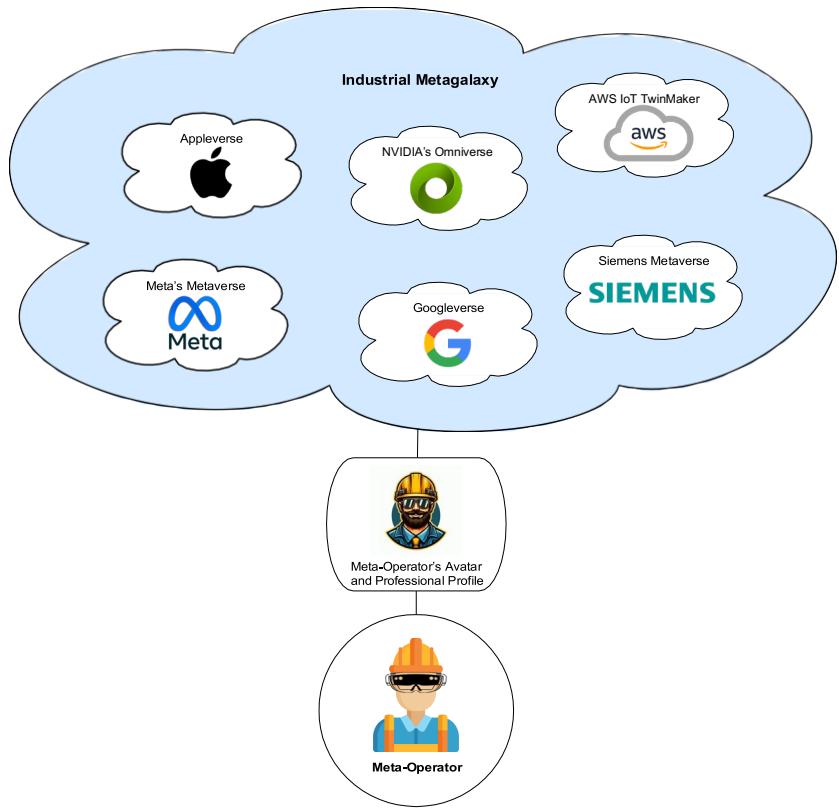

FIGURE 3: Example of an Industrial Metagalaxy.

TABLE 1: Comparison of the most relevant characteristics of traditional networks and the new metaverses.

|  | Traditional Internet | Traditional Industrial Networks | Commercial Metaverse | Industrial Metaverse |
|---|---|---|---|---|
| Main Functionality | Data exchange and processing | Physical equipment monitoring and control | 3D-content exchange and processing | Physical equipment monitoring and control through 3D interfaces |
| Application Field | Homes and companies | Manufacturing, product management or service delivery industries | Homes and companies | Manufacturing, product management or service delivery industries |
| Protocols and Standards | Unified protocols and standards | Many protocols and standards | Yet to be standardized | Yet to be standardized |
| Failure Severity | Low | High | Low | High |
| Required Reliability | Moderate | High | Moderate | High |
| Latency | 50 ms | 250 us - 10 ms | 1-10 ms | 250 us - 10ms |
| Required Determinism | Low | High | Low | High |
| Type of Network Traffic (in general) | Small packets of periodic or asynchronous traffic | Large packets of asynchronous traffic | Small and large packets of synchronous traffic | Small and large packets of synchronous traffic |
| Temporal Coherence | Non required | Required | Required | Required |
| Usual Operational Environment | Usually 'clean' environments where fragile equipment can operate | Hostile environments with dust, heat, noisy, vibrations and radio interference | Usually 'clean' environments where fragile equipment can operate | Usually 'clean' environments where fragile equipment can operate |
| Most Common Interaction Devices | Traditional computers, smartphones and tablets | Ruggedized computers, smartphones and tablets | VR headsets | AR and MR smart glasses |

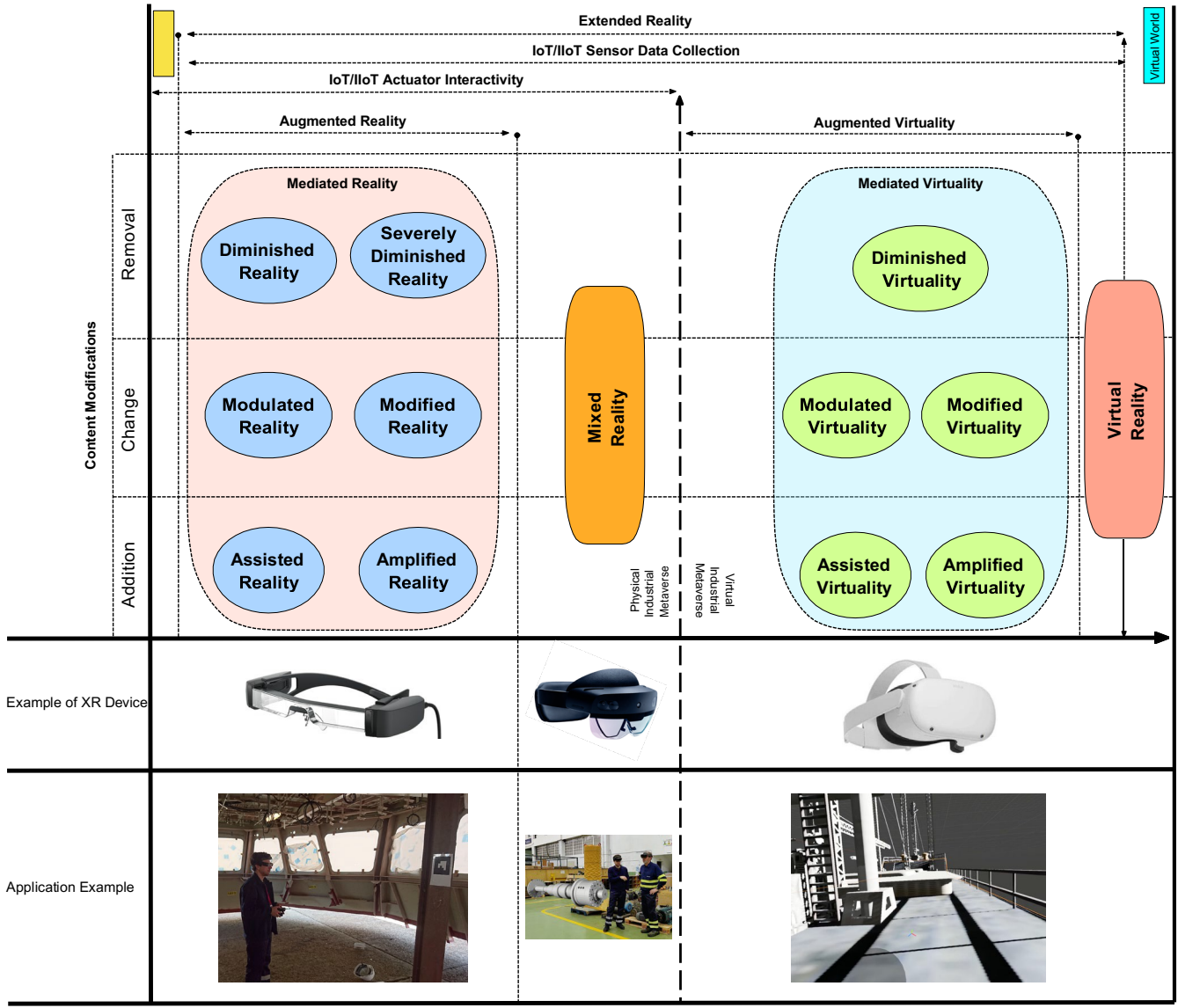

FIGURE 4: Main XR technologies for the Industrial Metaverse.

ditional information. The virtual content can be shown in simple scenarios (e.g., showing a warning when a specific time alarm is triggered, showing a notification received through a messaging app, depicting in real time the values of certain sensors) or it may require to first detect that certain situation has happened (e.g., detecting visually a surrounding object, determining that a Meta-Operator is in a specific scenario, detecting that the operator biometric patterns have reached a dangerous situation [59]).

- Amplified Reality (amR). It goes a step beyond aR and synchronizes the state of the provided additional information publicly, so all users can see the same content [60]. Thus, if a Meta-Operator changes the state of an object (i.e., he/she changes its properties) or adds certain content (e.g., adding notes on a task performed on a product [61]), all the users of the same Industrial Metaverse will perceive the same information.

- Modulated Reality (modR) and Modified Reality (MfR). They refer to similar concepts where the Meta-Operator's perception is altered through the filtering and modification of real elements. A simple example is a modR/MfR soldering mask that would lower the level of brightness coming from soldering in order to prevent damaging the Meta-Operator's sight. Another example would consist in adjusting digitally the shown content to a Meta-Operator depending on his/her sight prescription in order to avoid wearing regular prescription glasses under his/her AR device.
- Diminished Reality (DR). Type of AR that is able to hide or remove real elements from the Meta-Operator's perception. For instance, when a specific object needs to be located in a large industrial warehouse, it can be useful to remove from the sight of the Meta-Operator the elements that are not relevant for the search [62].
- Severely Diminished Reality (SDR). This kind of AR goes beyond DR by being able to remove the entirety of a real environment and even certain senses, thus resulting in a sort of sensory deprivation. For example, SDR can be useful in situations when Meta-Operators need to remain concentrated when performing a task, but other sensory inputs may distract him/her (e.g., noise coming from an industrial environment).

- Mixed Reality (MR). It presents virtual content overlaid with reality that not only can be seen, but it can also be interacted with and then produce changes in the real environment. Thus, surrounding objects react to the Meta-Operator's actions (e.g., hitting virtual machinery may damage it virtually) and are connected to reality. For instance, a virtual panel can be connected to an IIoT machine in order to control it without requiring to interact with its physical inputs [63].
- Augmented Virtuality (AV). It enhances a virtual environment with some information or elements of the real world. In this way, AV allows Meta-Operators to remain immersed in a virtual world but receive information from the real world. There are different kinds of AV that are analogous to the ones for AR but applied to a virtual world:
  - Assisted Virtuality (asV). It superimposes real-world content on the virtual world in order to provide additional and accurate information. For example, a Meta-Operator can be immersed in the digital twin of a factory and receive its actual performance data in real time.
  - Amplified Virtuality (amV). It enhances asV applications by synchronizing the state of the provided content among all users.
  - Modulated Virtuality (modV) and Modified Virtuality (MfV). These are types of AV that change the user perception on the surrounding virtual world. For instance, this can be useful in Industry 5.0 factories to ease a Meta-Operator's work: by filtering the shown virtual content, only the content required by the operator to perform his/her job is shown. MfV and modV can also be useful in certain Industrial Metaverses where roles are clearly defined (i.e., where the information to be shown to a technical operator differs from the one required by a corporate executive) or when the shown content depends on the Meta-Operator's age/experience (e.g., more experienced operators may require seeing certain raw information that is initially shown in a more user-friendly way to beginners).
    - Diminished Virtuality (DV). It is a type of AV that removes certain virtual elements from the Meta-Operator's perception. The use of DV in a regular public metaverse is straightforward when, for instance, having to prevent minors from watching restricted content. In the case of Industrial Metaverses, the uses of DV are similar to the ones provided by modV and MfV, since they are aimed at removing unnecessary objects from the sight of the Meta-Operator or at restricting the access to certain virtual content according to a role policy or to an internal hierarchy.
- Virtual Reality. It immerses Meta-Operators into an entirely artificial world. Thus, once entered into a VR-based Industrial Metaverse, it becomes a distinct reality that is detached from the real world. In industrial scenarios, for example, VR is really useful when delivering realistic simulations or when having to train operators in a specific skill.

It must be indicated that Figure 4, while it indicates that IoT/IIoT data can be collected by any XR technology, it only considers IoT/IIoT-device interactivity for AR and MR. This is due to the fact that, in an Industrial Metaverse, when a Meta-Operator needs to interact with machinery, for security reasons, he/she cannot remain in an entire virtual world. Nonetheless, note that it is perfectly possible to interact with IoT/IIoT devices when immersed in a virtual world and that an interaction may be very useful when developing asV and amV applications related to digital twins, since the shown information would come from real devices.

Moreover, it is worth noting that, due to the previously mentioned lack of IoT/IIoT interactivity, Figure 4 sets a limit between the physical and the virtual Industrial Metaverse. Thus, two different kinds of applications are distinguished in the Industrial Metaverse: the ones based on AR and MR that require interaction with real-world IoT/IIoT objects and the ones that occur in a virtual world.

In relation to Figure 4 it is also worth mentioning that at its bottom are included examples of XR devices (for AR, MR and VR) together with examples of industrial applications for each device: an aR application that makes use of markers to indicate to the Meta-Operator information on the tasks and materials to be used [13], an MR application for assembling clutches in a turbine workshop [64] and a VR application to train a ship crew in evacuations.

Finally, in relation to AR/MR devices, it is fair to indicate that their interaction with the surrounding environment depends on different technologies, which are summarized in Figure 5. In such a Figure it can be observed that AR/MR devices can rely on marker or markerless technologies. The former type makes use of fiducial markers (e.g., QR codes) or images to determine when to show certain information, or detect objects, faces or parts of a body (e.g., hand tracking). The latter type takes advantage of other mechanisms to determine the user location and his/her surrounding objects, like radiofrequency (RF) beacons (e.g., Bluetooth beacons [65]), GPS coordinates, motion/gesture detection techniques, spatial tracking [66], surface detection or world mapping [67].

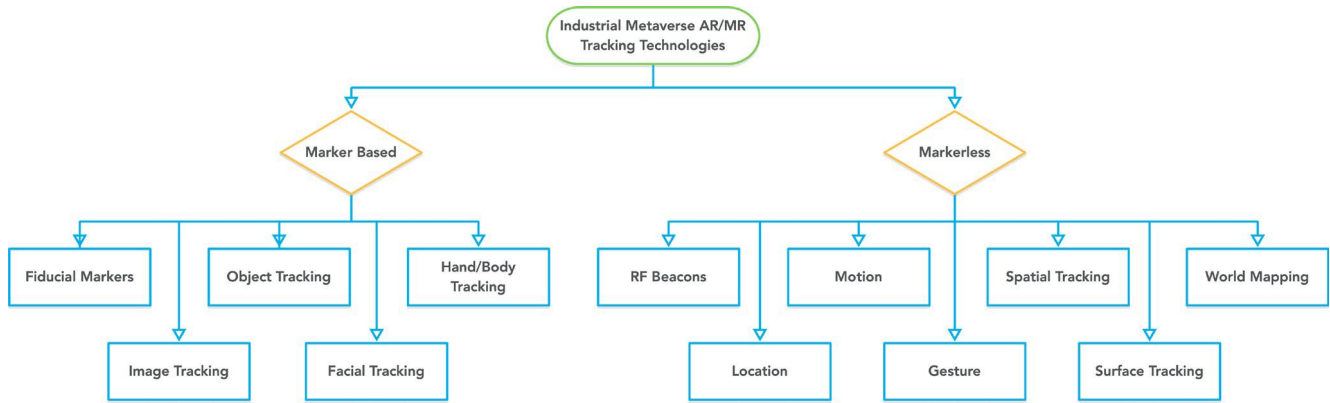

FIGURE 5: Main AR/MR Meta-Operator tracking technologies.

### D. OPPORTUNISTIC EDGE COMPUTING COMMUNICATIONS

Edge Computing is a paradigm that make it not necessary to send requests to remote clouds, which have multiple disadvantages that impact UX (i.e., relatively high delay), efficiency (e.g., high energy consumption), accessibility (access is restricted when maintenance tasks are performed, Denial-of-Service (DoS) attacks occur, when power outages happen or when Cloud servers become overloaded when a high number of users/devices access them concurrently) and privacy (e.g., public server exposure makes clouds prone to attacks related to data leaks) [68]. To tackle such issues, Edge Computing proposes a decentralized alternative where computing is performed by devices located at the edge of the network [69], [70]. Thus, edge devices can answer faster than a remote cloud and the amount of requests sent to the Cloud are reduced substantially [71].

Edge Computing has been already applied to Metaverse applications together with 6G communications and AI, which are also essential for the future of the Metaverse [72]. For instance, in [73] the authors present PolyVerse, a solution that makes use of locally-deployed Edge Computing devices that allows for the real-time projection of large virtual objects with a relatively low latency (with an average latency of 250 ms). Similarly, other researchers were able to reduce latency up to 50% respect to cloud-based Metaverse applications through the use of Edge Computing devices [74].

Opportunistic Edge Computing (OEC) systems can make use of Edge Computing devices to identify surrounding IoT/IIoT devices and offer them Edge Computing services opportunistically [20]. Given that IoT/IIoT devices are typically resource constrained and are dispersed across large environments, their connectivity and computational functions depend on external devices that may not be accessible constantly (commonly, a remote Cloud). In addition, such distributed IoT/IIoT devices are often powered by batteries, so it is important to reduce energy consumption and computational-resource utilization.

OEC systems are especially useful in situations where IoT/IIoT devices have no continuous Internet connectivity (e.g., when they are deployed in remote locations or when wireless communications are difficult, as it usually occurs in factories where there are many metallic objects [75]), when such devices have limited computing resources (e.g., processing power, internal storage) or when the deployed IoT/IIoT devices are static or have limited mobility, which prevents them from moving to places to communicate with other devices.

The collaborative nature of OEC solutions can be harnessed to tackle part of the limitations of traditional Cloud Computing based architectures, which traditionally have not been devised as energy-efficient solutions and they have scalability problems when dealing with massive IoT/IIoT implementations [68].

OEC has been recently become feasible and affordable thanks to the technological progress made on Single-Board Devices (SBCs), wearables and embedded IoT/IIoT devices, which are now able to provide enough computing power and reduced power consumption, thus being able to act as Industrial Metaverse gateways or smart end devices in Edge Computing architectures [76], [77].

OEC communications have similarities with the ones derived from the deployment of Mobile Ad-Hoc Networks (MANETS) [78]. However, MANETS focus on routing, while OEC goes beyond and provides Edge Computing services and additional communication capabilities [20]. In the literature several authors have previously proposed paradigms that are similar to OEC, but with diverse names like Proximal Mobile Edge Server [79], Mobile IoT [80] or Opportunistic Fog Computing [81], and has been applied to diverse IoT/IIoT applications for mobile communications [82], wildlife monitoring [83] or smart cities [84]. However, this latter kind of applications usually involve a relevant dependency: they rely on Internet connectivity to make use of the services provided by a remote Cloud.

As an example, Figure 6 shows an OEC architecture for an Industrial Metaverse Factory (called 'Meta-Factory'). In such an example, IIoT devices are deployed throughout the factory (IIoT devices A to F). Meta-Operators and autonomous vehicles (e.g., Automatic Guided Vehicles (AGVs), Unmanned Aerial Vehicles (UAVs) or other kinds of autonomous transport vehicles) monitor, interact and provide services to the deployed IIoT devices, which may be located in places where communications are not available or where it is not possible (or very expensive) to deploy communications infrastructure. Thus, when an IIoT device is detected by a Meta-Operator XR device (or by one of the autonomous vehicles), the former can make use of the OEC services provided by the latter. In addition, Meta-Operators can harness this kind of communications to collect data and interact with the deployed IIoT devices through Industrial Metaverse applications. Furthermore, when the services required by an IIoT device need to make use of Cloud services (e.g., when intensive computing is necessary), the hardware carried by Meta-Operators or by the factory autonomous vehicles can collect IIoT device requests and send them to the Cloud through the Routing Layer. Also, this way of operating allows for storing in the Cloud the data collected from the deployed IIoT devices and let them communicate with other remote IoT/IIoT networks or external services.

Finally, it is worth mentioning the main software components that an OEC system should implement in an Industrial Metaverse solution [20]:

- Peer discovery. In an Industrial Metaverse application it is essential for Meta-Operators to detect the surrounding IoT/IIoT devices and to establish a communication channel with other Meta-Operators. For such a purpose, it is necessary to implement a device discovery protocol that, ideally, should be secure and fast (Meta-Operators may be moving, so the communication window may be really narrow, so a fast device discovery protocol is needed). For instance, a node discovery protocol is described in [79].
- Peer routing. This service is related to the ability of routing the communications to/from a specific device, which usually requires to previously establish an efficient path to reach the destination.
- Data routing. It allows the OEC system to send information from one device to another when the receiving device is not within the communications range of the sending device.
- Resource sharing. This component is necessary to optimize resource use efficiency while delivering the necessary resources as close as possible to the IoT/IIoT and XR devices. Thus, response latency is reduced, which is essential for the UX of an Industrial Metaverse. For example, in [80] and [82] the authors deal with the issues that arise when implementing a resource-sharing service.

### E. STANDARDIZATION INITIATIVES

If the Metaverse wants to achieve a level of success similar to the one obtained by the Internet, standardization is necessary. Although standardization is still an ongoing effort, different organizations are carrying out initiatives for such a purpose (a good compilation of these initiatives can be found in [85]):

- ITU (International Telecommunication Union). The ITU established a Focus Group on Metaverse (FG-MV) that includes around 500 experts that work on the foundations of potential future standards [86]. Such a group has already delivered over 20 technical specifications and reports [87]. The mentioned documents define the concept of Metaverse [88], analyze the requirements for cross-platform interoperability [89] or detail the potential cyber-threats that can occur in a Metaverse [90].
- 3GPP. The 3rd Generation Partnership Project (3GPP) is well-known for its responsibility in the standardization of mobile telecommunications (e.g., GSM-2G, UMTS-3G, LTE-4G, 5G). The 3GPP is currently carrying out projects related to the Metaverse, like as a study on how to support tactile and multi-modality communication services (which includes a specific use case on a virtual factory) [91], a study on how to provide localized mobile Metaverse services [92] or another analysis on how to deliver XR services [93].
- Metaverse Standards Forum. Such a forum was presented as an organization to foster interoperability standards for an open metaverse, so that it will not require an intellectual property framework [94]. Thus, although the Metaverse Standards Forum is still in its beginnings, it has a significant number of operating groups that work in topics like 3D web interoperability, data asset management, interoperable characters/avatars or in network requirements and capabilities. Moreover, it has a specific working group in Industrial Metaverse interoperability. Thus, the output of the Metaverse Standards Forum it aimed at carrying out for the metaverse what 3GPP did for cellular networks [95].
- MPAI (Moving Picture, Audio, and Data Coding by Artificial Intelligence). It is currently performing different Metaverse-related activities, being the output of the most relevant the definition of the MPAI Metaverse model [98].
- MPEG (Moving Picture Experts Group). The MPEG has added to the MPEG-I suite the MPEG Immersive Video (MIV) standard, which has been designed to support XR applications with 6DoF visual interaction [96].
- IEEE. The IEEE has established a standard committee [97] that currently includes two working groups. One

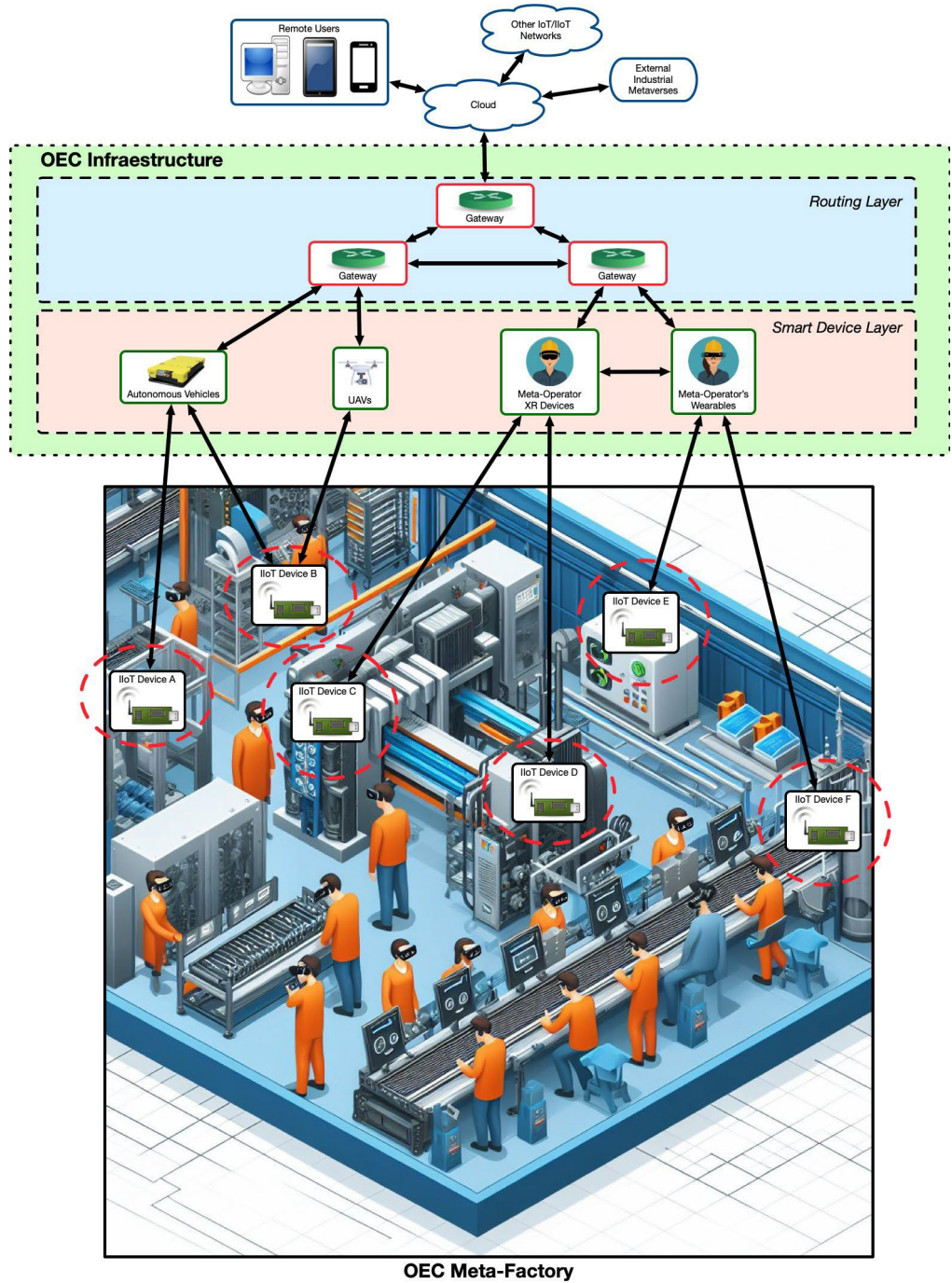

FIGURE 6: Example of OEC architecture for a Meta-Factory.

of them has been focused on AR for mobile devices, a topic that has been already the focus of the IEEE standard for AR learning experience models [99]. Moreover, the IEEE has established a working group for the IEEE P2048 standard, which is aimed at determining the terminology, definitions and taxonomy for the Metaverse [100], an another group for the IEEE P7016 standard [101], which seeks to create a methodology for developing Metaverses that consider the relevant ethical and social aspects. Furthermore, two IEEE initiatives have also been established: the Decentralized Metaverse Initiative [102] and the Persistent Computing for Metaverse Initiative [103]. The former is dedicated to the development and guidance for creating decentralized Metaverse, while the latter is focused on the technologies that are necessary to build, operate and upgrade the developed Metaverse experiences.

## III. FORGING A META-OPERATOR

### A. THE META-OPERATOR CONCEPT

A Meta-Operator can be defined as an industrial worker that follows the principles of the Industry 5.0 paradigm and interacts inside an Industrial Metaverse application essentially thanks to the use of a smart AR/MR device. Such a device embeds all the necessary technology to monitor the operator and the surrounding industrial environment.

As it was previously mentioned, for industrial environments, AR and MR are preferred to VR due to safety reasons, especially when having to interact with certain machines or with potentially dangerous processes. Although Meta-Operators can make use of VR headsets for certain Industrial Metaverse applications (e.g., training, purchases or 3D-content design), this article assumes that, in contrast to the Commercial Metaverse, AR and MR are the best XR technologies for working in industrial environments.

As it is illustrated in Figures 2 and 7, the AR/MR device is not only responsible for the AR/MR functionality, but also for acting as the operator gateway to the Industrial Metaverse, thus accessing data from multiple sources and interacting with local and remote IIoT devices, as well as with other Meta-Operators. In addition, as it is represented in Figure 8, each Meta-Operator's AR/MR device acts as a connectivity hub that is the center of the development of an Industrial Metaverse solution. Specifically, the core of the system is composed by the internal sensors, which often include an Inertial Measurement Unit (IMU), depth sensors and cameras that allow for tracking the operator movements and for creating a Wireless Body Area Network (WBAN) and/or a Wireless Personal Area Network (WPAN) to connect to external peripherals (for instance, Bluetooth devices like smart health wearables, access control systems or GPS receivers). The used AR/MR devices can also embed Wireless Local Area Network (WLAN) transceivers (e.g., IEEE 802.11 ac in the case of Microsoft HoloLens 2) to communicate with the multiple industrial Intranet software (e.g., IIoT platforms [104], digital twin software [105] or Industrial Cyber-Physical Systems (ICPSs) [106]), machines (e.g., industrial machinery, Edge Computing servers, a local cloud, Unmanned Aerial Vehicles (UAVs) [107]) and users (e.g., other Meta-Operators or workers that use a PC, a smartphone or a tablet) that connect to such a network. Furthermore, future XR devices will be able to communicate with Low-Power Wide Area Networks (LPWANs) like Sig-Fox or LoRaWAN [77] in order to collect data from sensors or to interact with remote actuators. Furthermore, the next generation of Wireless Wide Area Networks (WWANs) (e.g., 5G, 6G) will enable to connect Meta-Operators fast with remote users or servers in real time [108].

The primary characteristics of the main technologies used for establishing WPAN, WLAN, LPWAN, and WWAN networks are provided in Table 2.

### B. ADVANCED AR/MR GLASSES FOR A META-OPERATOR

A key component of a Meta-Operator is the carried AR/MR device, which is used for retrieving, collecting and visualizing data. Traditionally, AR/MR devices could be ruggedized for being used in industrial environments, but they were bulky and provided limited mobility (due to battery life or communications range). Luckily, Head-Mounted Displays (HMDs), tablets and smartphones have evolved substantially in recent years, allowing for increasing operator mobility when performing specific tasks. Nonetheless, tablets and smartphones require users to switch their attention between the performed task and the AR/MR application, which can be distracting. By enabling hands-free operation and overlapping reality and virtual content, AR/MR HMD devices eliminate the majority of such distractions, making them the most promising alternative for future Meta-Operators. Thus, Table 3 compares the most relevant characteristics of the latest and most popular HMDs. The following are the main key findings that can be obtained from the analysis of such a Table:

- Thanks to the last years price decrease, AR/MR devices can currently be purchased for a price similar to the one of industrial ruggedized smartphones and tablets. However, the most advanced HMDs are still expensive for massive deployments.
- AR/MR device features not necessarily depend on their price. For instance, Vuzix M400C were designed to visualize content from external PCs or Android smartphones, while currently there are cheaper HMDs (e.g., Meta Quest Pro) that provide powerful standalone environments.
- All the compared devices embed an IMU or some kind of indoor position tracking through integrated cameras, it is difficult to find AR/MR devices that provide a GPS for outdoor positioning.
- WiFi and Bluetooth transceivers are embedded into most AR/MR devices, so wireless connectivity at acceptable transfer rates are available.
- Most AR/MR devices rely on batteries, which usually provide between 2 and 6 hours of life.
- The technical specifications of the embedded display vary significantly among the compared models. Most AR/MR devices make use of see-through displays, but their field of view ranges from 16.8° to 110-120°, which, in terms of user experience, supposes a dramatic difference (the wider the field of view, the better the UX).
- The characteristics of the embedded cameras differ noticeably, ranging from 2 MP to 13 MP and from VGA to 4K video resolution.

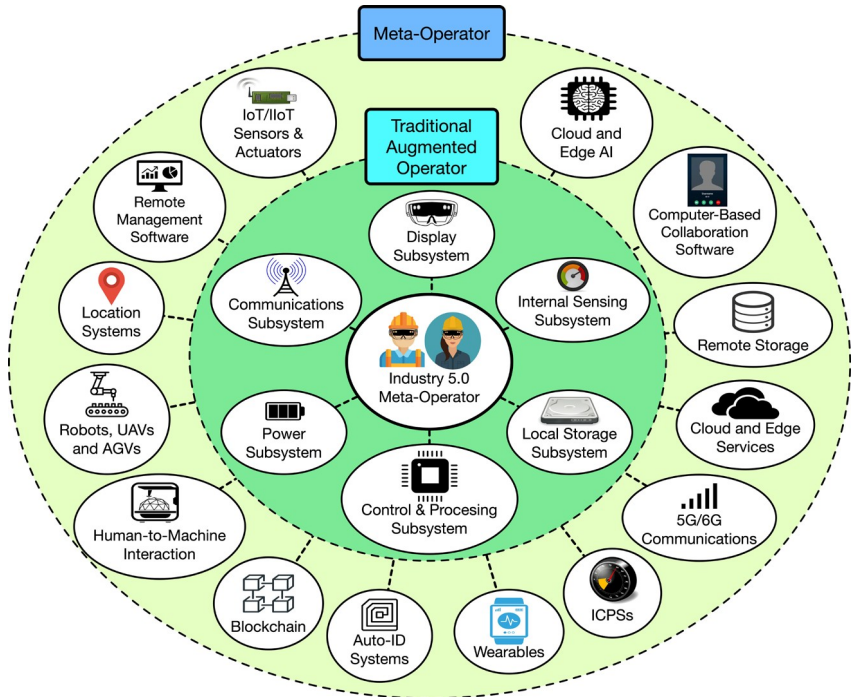

FIGURE 7: Comparison of the subsystems of a traditional AR/MR-based operator versus a Meta-Operator.

TABLE 2: Main characteristics of the most popular communications technologies for Meta-Operators.

| Type | Technology | Frequency Band | Typical Maximum Range | Data rate | Main Features |
|---|---|---|---|---|---|
| WBAN/WPAN | UWB | 3.1-10.6 GHz | Short-range | Up to 480 Mbps | Precise ranging, high data rates |
| | ANT+ | 2.4 GHz | 30 m | 20 kbit/s | Very low power consumption, up to 65,533 nodes |
| | Z-Wave | 800-900 MHz | Up to 100 meters | Up to 100 kbps | Low power, home automation and IoT applications |
| | HF RFID | 3–30 MHz (13.56 MHz) | a few meters | <640 kbit/s | Low cost, in general it needs no batteries |
| | LF RFID | 30–300 KHz (125 KHz) | <10 cm | <640 kbit/s | Low cost, it needs no batteries |
| | NFC | 13.56 MHz | <20 cm | 424 kbit/s | Low cost, it needs no batteries |
| | UHF RFID | 30 MHz–3 GHz | tens of meters | <640 kbit/s | Low cost, it usually needs no batteries |
| | WirelessHART | 2.4 GHz | <10 m | 250 kbit/s | Compatibility with the HART protocol |
| WLAN | ZigBee | 868-915 MHz, 2.4 GHz | <100 m | 20−250 kbit/s | Very low power (batteries last months to years), up to 65,536 nodes |
| | Bluetooth 5 LE | 2.4 GHz | <400 m | 1,360 kbit/s | Low power (batteries last from days to weeks) |
| | IEEE 802.11b (Wi-Fi 1) | 2.4 GHz | 35 meters (indoor), 140 meters (outdoor) | Up to 11 Mbit/s | First widely adopted WLAN standard |
| | IEEE 802.11a (Wi-Fi 2) | 5 GHz | 35 meters (indoor), 120 meters (outdoor) | Up to 54 Mbit/s | Uses a different frequency band to avoid interference |
| | IEEE 802.11g (Wi-Fi 3) | 2.4 GHz | 38 meters (indoor), 140 meters (outdoor) | Up to 54 Mbit/s | Backward compatible with 802.11b |
| | IEEE 802.11n (Wi-Fi 4) | 2.4 GHz and 5 GHz | 70 meters (indoor), 250 meters (outdoor) | Up to 600 Mbit/s | Introduced MIMO |
| | IEEE 802.11ac (Wi-Fi 5) | 5 GHz | 35 meters (indoor), 250 meters (outdoor) | Up to 3466.7 Mbit/s | Introduced MU-MIMO |
| | IEEE 802.11ax (Wi-Fi 6) | 2.4 GHz and 5 GHz | 30 meters (indoor), 90 meters (outdoor) | Up to 9608 Mbit/s | Improved efficiency, capacity, and performance in dense environments |
| | IEEE 802.11be (Wi-Fi 7) | 2.4 GHz, 5 GHz, and 6 GHz | TBD | Up to 46,120 Mbit/s | Extremely High Throughput. Improved spectral efficiency, latency, and power consumption |
| LPWAN | DASH7/ISO 18000-7 | 315–915 MHz | <10 km | 27.8 kbit/s | Very low power (batteries last from months to years) |
| | IQRF | 868 MHz | hundreds of meters | 100 kbit/s | Low power, long range |
| | SigFox | 868-902 MHz | 50 km | 100 kbit/s | It makes uses of private cellular networks |
| | Weightless-P | License-exempt sub-GHz | 15 Km | 100 kbit/s | Low power |
| | LTE-M (LTE for Machines) | Licensed LTE spectrum | 1-10 km | 1 Mbps | Directly integrated into LTE and 5G networks, supports voice and mobility |
| | Ingenu (RPMA) | 2.4 GHz | Up to 80 km | Up to 625 kbps | Random Phase Multiple Access, long-range, low power |
| | MiWi | 2.4 GHz | Up to 100 meters | Up to 250 kbps | Suitable for short-range, low-power applications |
| WWAN | NB-IoT | LTE frequencies | <35 km | <250 kbit/s | Low power, long range |
| | 4G LTE | 700 MHz - 2600 MHz | <35 km | <100 Mbit/s | High speed, wide coverage |
| | 5G | mmWave frequencies (24 - 100 GHz) | <3 km | <10 Gbit/s | High speed, wide coverage, low latency |
| | WiMAX | 2.5-5.8 GHz | Up to 50 km | Up to 1 Gbps | Broadband wireless access, fixed and mobile connectivity |
| | 6G | Sub-THz frequencies | <1 km | <100 Gbit/s | High speed, wide coverage, low latency, AI integration |

TABLE 3: Main features of potential AR/MR devices for Meta-Operators.

| Model | Weight | Price | Main Specs | Sensors | Communications | Battery | Display | Input/Output | Software |
|---|---|---|---|---|---|---|---|---|---|
| Apple Vision Pro [109] | ~500 g | $ 3,499 | Apple M2, Apple R1 | Currently unknown, but the device provides 6 Degrees of Freedom (DoF) with device tracking (probably through an IMU) and hand/gesture tracking through 12 built-in cameras and a LiDAR. | WiFi, Bluetooth | Battery life: up to 2 hours of use (optional external power bank) | Stereoscopic dual ~3400x3400 see-through micro-OLED displays at 90Hz, approximate field of view between 100° and 120° | 12 integrated cameras, 6 microphones, stereo speakers | Based on visionOS |
| ATHEER ThirdEye Gen's X2 MR Glasses [110] | ~170 g | $ 2,450 | Integrated CPU/GPU, 4 GB of RAM, 64 GB of storage | 9-DoF IMU with accelerometer, gyroscope and magnetometer, ambient light sensor, thermal sensor | WiFi, Bluetooth, GPS | 1,750 mAh battery | Stereoscopic dual 720p see-through displays at 60 fps with a field of view of 42° | 13 MP HD camera, noise-cancelling microphone, 2 additional wide angle cameras, diverse control mechanisms (i.e., gaze, voice and gesture control, wireless controller) | Based on Android 8.1, VisionEye SDK, support for Unity |
| Epson Moverio BT-45C [111] | 545 g (with headband) | $ 1,499 | Qualcomm Snapdragon XR1, 4 GB of RAM, 64 GB of storage | Compass, gyroscope, accelerometer, light sensor | Bluetooth 5.0 | 3,400 mAh battery | Dual 1080p 60 Hz 0.453" Si-OLED panels with a field of view of 34° (diagonal) | 8 MP camera, USB Type-C 2.0 with DP, 3.5 mm audio jack | Supports Windows 10/11 or Android 8.0 or later host devices |
| HTC VIVE XR Elite [112] | 625 g | ~$1,099 | Qualcomm Snapdragon XR2, 12 GB of RAM, 128 GB of storage | It gets 6 DoF through 4 integrated cameras, depth sensor, g-sensor, gyroscope, proximity sensor | WiFi 6E, Bluetooth 5.2, USB type-C | Removable and hot-swappable 24.32 Wh battery that lasts up to 2 hours | Stereo see-through LCD display with 1920x1920 pixels per eye at 90Hz with a Field of View (FoV) of 110° | 16 MP RGB camera, 4 tracking cameras, stereo speakers, two microphones | Based on Android with OpenXR support for Unity and Unreal Engine |
| Meta Quest Pro [113] | 722 g | ~$999 | Qualcomm Snapdragon XR2+, 12 GB of RAM, 256 GB of storage | It gets 6 DoF through 5 integrated cameras, 5-IR eye/face tracking sensors | WiFi 6E, Bluetooth 5.2, USB type-C | Battery life: 2-3 hours depending on use | See-through QLED display with 1800x1920 pixels per eye at 90Hz with a FoV of 106° | 5 external and 5 internal cameras (external for hand tracking, internal for eye/face tracking), controllers also include embedded cameras, four speakers, three microphones | Based on Android: Meta Presence Platform; support for Unity, Unreal Engine and OpenXR; with multiple APIs and SDKs (e.g., Interaction SDK, Voice SDK, Movement SDK, Meta Avatars SDK). |
| Microsoft HoloLens 2 [56] | 566 g | ~$3,500 | Second-generation custom-built holographic processing unit, Qualcomm Snapdragon 850 compute platform, 4 GB of RAM, 64 GB of storage | IMU (accelerometer, gyroscope, magnetometer), 4 visible light cameras (for head tracking), 2 infrared cameras (for eye tracking), 1 MP time-of-flight depth sensor | WiFi (IEEE 802.11ac 2x2), Bluetooth 5, USB type-C | Battery life: 2-3 hours with active use | See-through holographic lenses (waveguides), 2k 3:2 light engines, >2.5k radiants (light points per radian), Field of view: 52° | 8 MP photos, 1080p30 video, built-in spatial sound, 5-channel microphone array | Windows 10 with Windows Store, human tracking: hand, eye and voice tracking (with iris recognition based security) |
| Vuzix Blade 2 [114] | ~90 g | ~$1,380 | Quad-core ARM, 40 GB for storage | Gyroscope, accelerometer, magnetometer | WiFi, Bluetooth | n/a (the device is powered through a smartphone, tablet or computer) | 480p 1:1 aspect ratio color display with 2000 nits of brightness and 20° field of view | USB 2.0 Micro-B port, 2-axis touchpad with multitouch support, 8MP camera and integrated barcode scanner | Based on Android 11 |
| Vuzix M400C [115] | ~ 85 g | $ 1,299 | Monocular display (computer-powered AR device) | Gyroscope, accelerometer, magnetometer | USB 3.1 (USB-C that provides DisplayPort) | n/a (computer-powered device) | 16:9 OLED display with 2000 nits peak brightness and 16.8° field of view | 13 MP camera, LED flash, and barcode scanner. 4 control buttons, voice control and 2-axis touchpad with multi-touch support. | Compatible with host devices that support DisplayPort output over USB Type-C |
| Magic Leap 2 [116] | ~ 260 g with headstrap | $3,299 with controllers | AMD Quad-core Zen2 x86 CPU, AMD RDNA 2 GFX10.2 GPU, 16 GB of RAM, 256GB of internal storage | Accelerometer, gyroscope, magnetometer, altimeter, ambient light sensor, 4 gaze tracking cameras, depth sensing camera, RGB camera and 3 wide-FoV scene-tracking cameras | WiFi (IEEE 802.11 ax), Bluetooth 5.1 LE | Rechargeable battery with up to 3.5 hours of use | 1440x1760 resolution 120 Hz 2000 nits of brightness displays with a 70° field of view | 12.6 MP camera with 4K 30 FPS video recording, voice assistant, hand and gesture tracking, wireless touch sensitive controller, speakers with spatial audio | Magic Leap OS, based on AOSP (Android Open Source Project) |

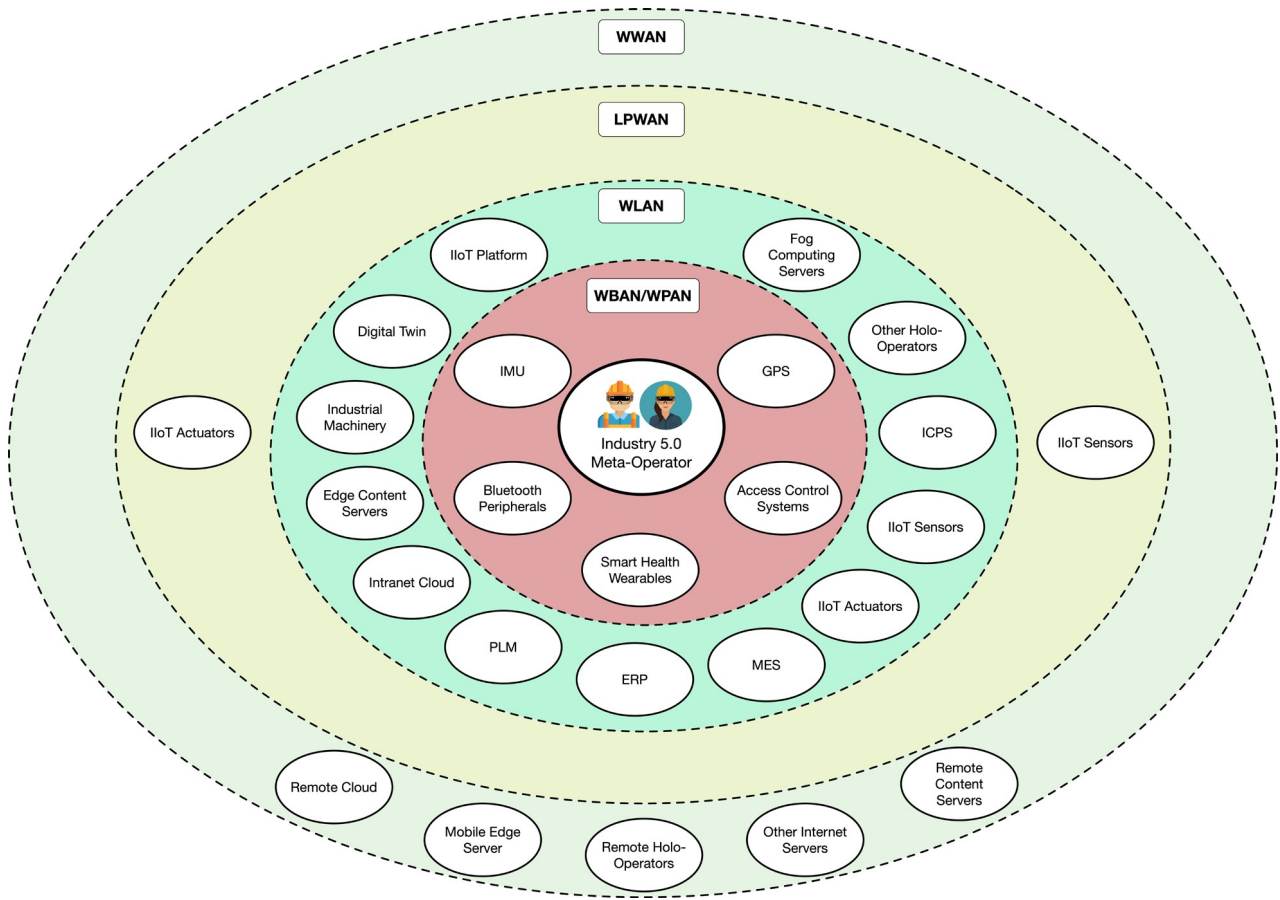

FIGURE 8: Connectivity of a Meta-Operator.

### C. USEFUL ACCESSORIES FOR INDUSTRIAL METAVERSE APPLICATIONS

Most of the latest AR/MR devices are able to track the hands and gestures of Meta-Operators to interact with the displayed virtual content, but there exist specific HMI devices that make industrial tasks more precise and agile.

The following are some of the most popular AR/MR accessories for Industry 5.0 scenarios:

- *i* Traditional controllers. Like gamepads, traditional AR/MR controllers provide buttons and arrows to provide interactivity with the content in an agile manner. For instance, Microsoft HoloLens is able to make use of Microsoft's Clicker, which removes the need for using hand tracking to interact with virtual content, so the Meta-Operator only has to look at the virtual object that he/she wants to interact with and then press the clicker. There are more sophisticated controllers, like the Meta Quest Touch Pro controllers, which are a pair of controllers for Meta Quest headsets that integrate 3 cameras per controller to map their position and rotation in real time. Such a way of operating allows them to be used close to a paired headset to avoid losing their tracking capabilities. In addition, Meta Quest Touch Pro Controllers include a removable stylus that enables them to be used like a pen.
- *i* Interaction gloves. Gloves are useful for situations when hand tracking is not available through the Meta-Operator HMD or when additional feedback is necessary to be provided [117]. For instance, Teslaglove [118] is a force-feedback haptic glove for VR applications that embeds per-finger force feedback motors that change their resistance during actions like grabbing objects in an Industrial Metaverse. The Teslaglove goes even further: it includes sensors and actuators to increase the sensation of touch, to track finger positions with very high accuracy and to monitor the heart rate and blood oxygen levels of the Meta-Operator. Another example of interaction glove is the Diver-X Contact Glove [119], which is a tracking and haptic glove able to track finger flexing and thumb position.
- *i* Haptic devices. Haptic feedback has been traditionally used in gaming, but it can go to a whole new level when applied inside a Metaverse. For example, the bHaptics TactSuit X16 [120] is a haptic vest that makes use of 16 haptic motors that allow users to feel different physical sensations. Another haptic vest is the Woojer VestEdge [121], which processes the received audio to generate haptic sensations on the back, sides and chest of the user.
- Full-body tracking. There are accessories able to estimate the Meta-Operator body position. Some of such accessories are just small wearables that can be carried by the Meta-Operator. For instance, Sony Mocopi [122] is a set of wearable tags that use 3-DoF sensors and Machine Learning (ML) models to track the user body position. If more precision is required, there exist full-body tracking solutions like HaritoraX [123], which makes use of 9-axis IMUs that are attached to different parts of the user body to estimate the position of his/her limbs.
- Face tracking. When VR-based experiences are required in an Industrial Metaverse and social interactions are performed, facial gestures are really useful to enhance UX. For such a purpose, there are device like Vive Focus 3 Facial Tracker [124] or HTC Vive Facial Tracker [125], which can capture facial expressions and mouth movements in real time.
- Eye tracking. Although the latest AR/MR devices already include gaze and eye tracking systems, it is possible to carry out eye tracking by making use of external devices like Vive Focus 3 Eye Tracker [126], which is a specific eye tracking module that provides independent tracking for each eye through two cameras and an infrared lighting system.
- Feet tracking. Monitoring the position of a Meta-Operator's feet is really useful in immersive Metaverse environments that require to determine with accuracy the user position. For instance, Surplex is a pair of shoes that embed IMUs and pressure sensors to estimate a Meta-Operator position without deploying a 6-DoF camera-based tracking system. Such an independence from using cameras implies that the system can operate out of the line of sight of such cameras or from base stations. In addition, avoiding the use of cameras prevents potential interference created by occlusions, which usually end up impacting the tracking process. Another examples of foot tracking system are Cybershoes [127], which are a special pair of shoes that are worn over the user's shoes and that are able to monitor walking motion and rotations. Moreover, Cybershoes can be used together with a special chair (called Cyberchair) and a carpet (Cybercarpet) that jointly can determine if a user is sitting properly or that is walking with a specific amount of friction.
- Omnidirectional treadmills. For VR-based metaverses, the amount of physical space for the users is essential when walking through large virtual scenarios. To avoid such a limitation, omnidirectional treadmills allow users to move freely but over a kind of treadmill that simulates the actual movement. For instance, Virtuix Omni One [128] is an omnidirectional treadmill based on an articulated arm that supports the user, providing 360° movements, including walking, running or kneeling. A similar treadmill is the KAT Walk C2 [129], but it incorporates specific shoes with optical sensors that allow for determining the position of the user feet accurately.

### D. INDUSTRIAL METAVERSE SOFTWARE PLATFORMS

Some examples of Industrial Metaverse software platforms along with their key characteristics can be seen in Table 4. The primary software platforms for Industrial Metaverse, which are quite recent, concentrate on empowering enterprises (e.g., real-world machines, factories). They achieve this by seamlessly integrating the physical and digital realms, with a particular emphasis on sustainable practices and accelerated operations. Some of the platforms envision an industrial metaverse that goes beyond and it accurately reflects transportation systems or even cities.

Leading companies in this field, such as Nokia, Siemens, and Microsoft, are making significant strides. While Unity, Microsoft and Meta's influence extends beyond industry-specific applications, they play a pivotal role in shaping the metaverse landscape, contributing to its growth and evolution.

In addition to Extended Reality (XR) and digital twins, blockchain technology is also having a profound impact on the overall ecosystem [132], [133].

### E. INTEGRATION OF AR/MR WITH IOT/IIOT

AR/MR technologies are currently considered as one of the most promising interfaces for visualizing and exploring the large complex information provided by IoT/IIoT applications. As of writing, no significant mature developments on such an integration can be found on the literature, but a few authors have proposed interesting alternatives. For instance, in [63] a framework is proposed to integrate AR/MR devices with IoT solutions through widely used standard communication protocols and open-source tools. In addition, other researchers have contributed to the field with demonstrators. For instance, in [151] the authors present a Proof-of-Concept (PoC) that monitors metal shelves with strain gauges and that has a QR code attached. When the operator scans the QR code, identification data are sent to a cloud and a simulation model designed with Matlab provides a stress analysis that is visualized through a pair of F4 smart glasses.

Other authors focused on enabling automatic discovery and relational localization to build contextual information on sensor data [152]. In the case of the work detailed in [153], the researchers describe a scalable AR framework that acts as an extension of the deployed IoT infrastructure. In such a system, recognition and tracking information is distributed over and communicated by the objects themselves. The tracking method can be chosen depending on the context and is detected automatically by the IoT infrastructure. The target objects are filtered by their proximity to the user.

In contrast, in [154] the authors make use of HoloLens smart glasses, which integrate Mobius, an open-source OneM2M IoT platform. However, in such a work the authors considered that further work will be needed in order to consider the various requirements defined by OneM2M.

TABLE 4: Key characteristics of main Industrial Metaverse software platforms.

| Project/Platform | Release Date | Objective | Company | Domain |
|---|---|---|---|---|
| Activision Blizzard [130] | 2022 | Virtual worlds for gamers. Developers of games such as Call of Duty ®, World of Warcraft ®, Diablo ® and Candy Crush ™. | Acquired by Microsoft in October 2023 [131] | AR and VR |
| Altspace VR [134] | Established in 2013 and began commercializing its products in 2015. | The platform was primarily made up of user-generated worlds. It hosted a wide variety of virtual events (e.g., fashion show for 2020 Paris Fashion week). | Redwood City, California, USA. Acquired by Microsoft in 2017. Microsoft shut down AltspaceVR on March 10th, 2023, shifting its focus to support immersive experiences powered by Microsoft Mesh. | VR |
| Aria [135] | September 2020 | Project Aria is a research device from Meta, worn like a regular pair of glasses. It enables researchers to accelerate the study of AR and AI from a human perspective. | Meta | AR |
| Axie Infinity [136] | May 2018 (Breeding game released) | Virtual world filled with creatures known as Axies (Pokemon-inspired creatures) that can be battled, bred, collected and traded on an open market place. | Sky Mavis (Ho Chi Minh City, Vietnam) | Entertainment. Web3, VR, NFT and blockchain. |
| Battle infinity [137] | 2022 | Gaming ecosystem that hosts multiple Play-to-Earn battle games integrated with a Metaverse world called 'The Battle Arena'. | Yuga Labs | Cryptocurrency, AR, and VR |
| Bloktopia [138] | March 2022 | Metaverse with an unprecedented VR experience for the crypto community. A decentralized VR skyscraper made of 21 floors. The number 21 was chosen in honor of Bitcoin's total supply of 21 million, programmed by its inventor Satoshi Nakamoto. | Bloktopia, Isle of Man, Ireland | Multiple contexts, cryptocurrency, blockchain, VR |
| Chetu Metaverse Development Services [139] | - | Software solution providers or different industries. | Chetu | VR, AR, NFT, blockchain. |
| Decentraland [140] | February 2020 | Decentralized virtual reality platform that allows users to create, explore, and interact in a virtual world. | Decentraland Foundation | VR, NFT, blockchain. |
| Gala Games [141] | 2019 | Platform that brings together a community of gamers and NFT enthusiasts. Currently developing the VOXverse | Blockchain Game Partners, Inc., also known as, Gala Games, Gala Music and Gala Film. | Entertainment |
| Horizon Workrooms [142] | 2021 | Immersive virtual office where you can meet teammates, brainstorm ideas or share presentations, whether you are wearing a Meta Quest headset or joining from a 2D screen. | Meta | AR and VR |
| Industrial Metaverse [143] | 2024 (announced in CES 2024) | Siemens Xcelerator portfolio of industry software with Sony's new spatial content creation system | Siemens partners with Sony, AWS, Red Bull Racing, Unlimited Tomorrow, and Blendhub | Manufacturing |
| Infosys metaverse foundry [144] | 2022 | It aims to accelerate and enhance enterprises' virtual-physical interconnections. It helps enterprises navigate the metaverse by partnering with them through the Discover-Create-Scale cycle. | Infosys | Examples: smart farming, virtual living labs, VR stores, life-like avatar powered collaborative experiences. |
| Microsoft Mesh [145] | 2021 | Significant part of Microsoft's industrial metaverse strategy including avatars and immersive 3D spaces (e.g., Teams). | Microsoft | AR and VR |
| NAKAverse [146] | 2021 | Metaverse to build the city of dreams on its blockchain-powered multi-faceted platform. | Nakamoto Games (Web3 gaming ecosystem) | Entertainment. VR and blockchain. |
| Nokia Industrial Metaverse [147] | - | It provides the foundation necessary to harness the potential of the metaverse leveraging its expertise in critical networks, connectivity, and digital transformation. A recent study conducted jointly by EY revealed that companies already deploying industrial metaverse use cases are experiencing significant benefits. | Nokia | Industrial and enterprise metaverses. |
| NVIDIA Omniverse platform [148] | 2020 | It provides developers with the building blocks–developer tools, APIs and microservices–to bridge data silos, connect teams in real time, and create physically accurate world-scale simulations, powered by OpenUSD and NVIDIA RTX. | 2020 | Standard, enterprise and cloud metaverses. |
| Unity [149] road to the metaverse: on-demand sessions | - | Foster knowledge, promote education. Build a strong foundation and develop skills with Unity experts during masterclass sessions. | Unity | Real-time 3D, AR, VR, digital twin, IoT and collaborative multi-user experiences. |
| Voxels [150] (formerly Criptovoxels) | 2018 | Free to play, browser based metaverse. Artists and builders can buy land and create. | Voxel Inc. | VR, NFT, blockchain. |

Microsoft HoloLens smart glasses are also used in [155] to enhance health and safety monitoring in construction sites. Specifically, the authors integrate information from IoT sensors and notifications (e.g., alerts, expiration dates) into a Message Queuing Telemetry Transport (MQTT) based platform. Such notifications are then collected by the smart glasses in order to show them to the construction operators to prevent safety issues. Besides MQTT, other protocols can help to integrate IoT/IIoT devices with Industrial Metaverse XR devices, like HTTP, OPC-UA, CoAP, AMQP, XMPP or MODBUS-TCP. A detailed description of such protocols is out of the scope of this article, but the interested reader can find further information in [156] and [157].

*F. INDUSTRIAL DIGITAL TWINS*

The concept of Digital Twin is inherently related to the Industrial Metaverse and to IIoT [158], [159], since it involves creating digital synchronized replicas of physical devices (e.g., machinery, sensors, actuators), materials [160], environments (e.g., assembly lines, factories [161]), systems (e.g., ICPSs) or people (e.g., consumers [162], Meta-Operators).

A key characteristic required by a Digital Twin is its connectivity, since they need to interconnect the physical and digital world. Thus, physical assets have to incorporate sensors and/or actuators that exchange data with digital systems. In addition, it is interesting that Digital Twins include traceability (to analyze their actions after experiencing a failure), re-programmability (to create new versions of the initial digital asset) and modularity (to customize the different submodules depending on the industry/application) [163]. Furthermore, Digital Twins can operate in an isolated manner or create more complex systems through Digital Twin Networks [164].

Digital twins have previously been used for implementing multiple industrial applications. For example, Renault presented in 2022 the digital twin of a vehicle [165]. Moreover, in [166] the authors describe the multiple applications related to a Healthcare Metaverse and propose the use of digital twins of the patients. Another example is presented in [167], where the authors detail the design and implementation of a flexible manufacturing system, while in [168] the digital twin of a production cell is created for educational purposes. Many more examples and applications, as well as a thorough review on the topic of Digital Twins can be found in [169].

## IV. DESIGNING AR/MR APPLICATIONS FOR THE INDUSTRY 5.0 METAVERSE
### A. COMMUNICATIONS ARCHITECTURE

Industrial Metaverse applications have been traditionally deployed in a communications architectures like the one depicted in Figure 9, which is divided into three main layers:

- *ı̇* XR-device layer. It is composed by traditional AR/MR/VR devices that run Industrial Metaverse applications. Two scenarios are distinguished: static (e.g., factories, workshops, offices) and dynamic (e.g., when operators have to be on the move, traveling in vehicles or walking through large environments that can change dynamically like construction sites or assembly lines). Besides traditional XR devices like HMDs, tablets and smartphones, in the case of static scenarios, projectors can be deployed, thus enabling providing Spatial AR (SAR) applications [170].
- *ı̇* Data-routing layer. It routes the data exchanged between the deployed XR devices and the industrial cloud. It is essentially made of routing infrastructure like gateways or wireless routers.
- *ı̇* Industrial cloud. It is usually a server farm where the industrial applications are executed, including the ones related to a company Industrial Metaverse or Metagalaxy. Such Metaverse-related deployments receive information and content from other industrial software, like SCADA (Supervisory Control And Data Acquisition), Product Lifecycle Management (PLM), Enterprise Resource Planning (ERP), CAD, IIoT or Manufacturing-Execution System (MES) software.

Although the architecture depicted in Figure 9 has been successfully deployed in many industrial scenarios, it involves three main limitations:

- *ı̇* Scalability. Although a server farm can be scaled in proportion to the number of expected Meta-Operators that will interact with it, the practical implementation of such a scalability is not straightforward or cheap, so it is necessary to devise novel architectures [171]. The main problem is that the Cloud can become a bottleneck when processing large amounts of requests from the XR devices.
- *ı̇* Latency. In addition to a potential cloud saturation, the fact of having to interact with a remote cloud implies that network latency is higher than when using local devices.
- *ı̇* Single point of failure. In situations when an Industrial Metaverse depends on a Cloud, if such a Cloud stops operating properly (e.g., power outages, communication problems, cyberattacks), the whole system stops working.

To tackle such previous problems, it is possible to harness the latest technologies and to provide complex Metaverse services [172], as well as to create the basis for deploying advanced Meta-Operating systems [173]. Thus, an example of advance architecture for implementing Industrial Metaverses is shown in Figure 10. In such a Figure there are five main layers:

- *ı̇* XR-Device Layer. It is like the one previously described for a traditional architecture, but it is able to communicates directly with the surrounding IoT/IIoT devices and with the OEC Layer.
- *ı̇* IIoT Device Layer. It is composed by the deployed IIoT devices.

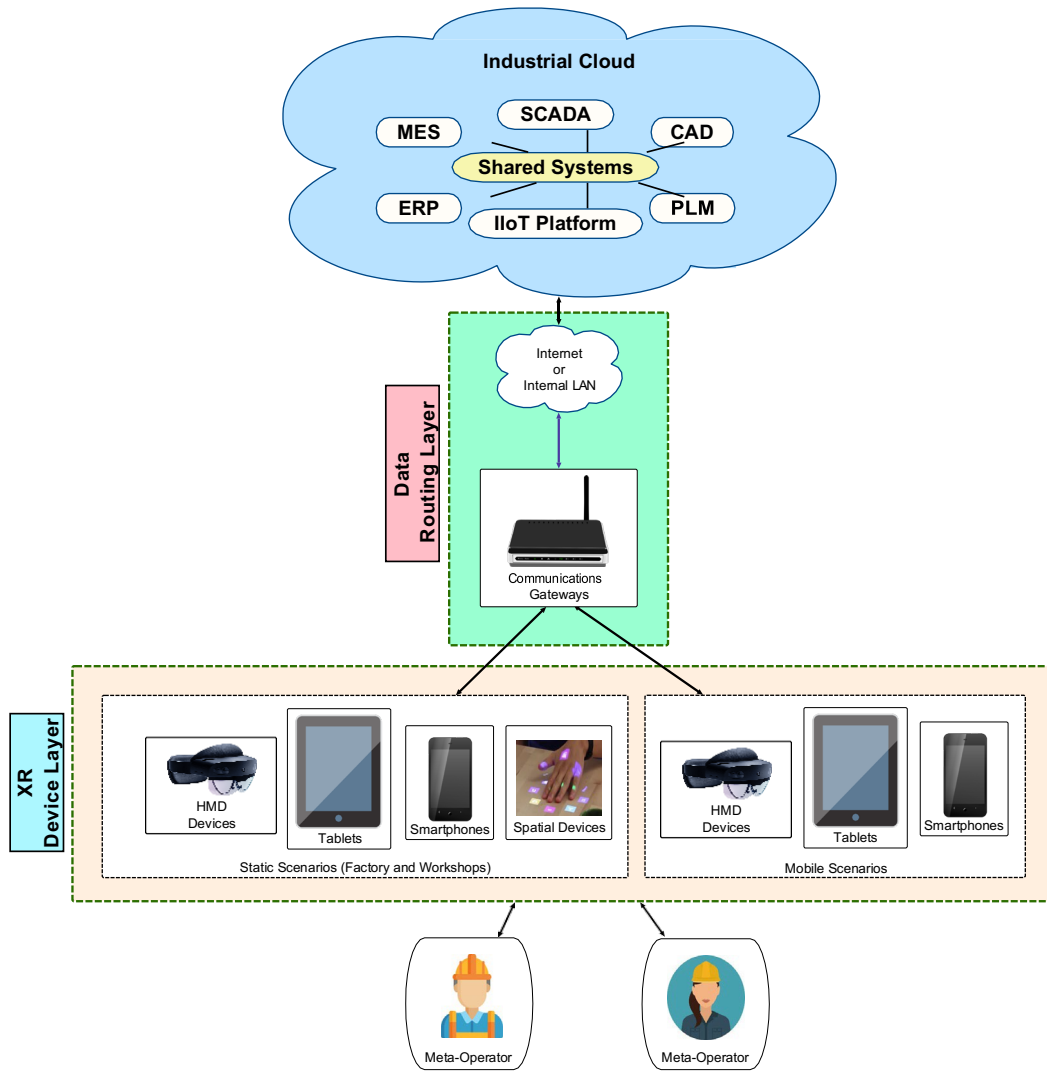

FIGURE 9: Traditional architecture for providing Industrial Metaverse applications.

- OEC Layer. It is composed by OEC devices like simple SBCs (e.g., fog computing gateways), powerful computers (e.g., Cloudlets) or Mobile-Edge Computing (MEC) nodes [69], [71], [174]. Such OEC devices exchange data opportunistically with the hardware carried by the Meta-Operators and with the surrounding IIoT devices. Moreover, OEC devices can exchange data among them so as to provide more complex and location-aware services.
- Industrial Cloud. It stores data, provides services to remote users and responds to the requests received from OEC nodes.
- Industrial Metagalaxy. They enable remote Meta-Operators to get access to external metaverses that can be useful in certain industries. For instance, Meta-Operators can join metaverses aimed at providing remote healthcare assistance or remote-training, or for purchasing goods.

### B. SOFTWARE ARCHITECTURE

Together with the communications architecture, it is necessary to define a software architecture that, ideally, should be composed by independent modules that can be replaced easily. Thus, such software modules can be decoupled from the rest of the software components of the architecture and then they can be integrated into third-party projects easily.

The main components that should be part of an Industrial Metaverse application and their interconnections are shown in Figure 11. At the bottom of such a Figure, on the left, is the XR subsystem, which includes the software components that are executed by XR devices. Among such components is the Interaction Manager module, which is in charge of detecting and handling user inputs (e.g., hand movements, gestures, external controller inputs). The IoT/IIoT Application Programming Interface (API) is responsible for the bidirectional communication with the IoT/IIoT devices, while the World Sharing service is in charge of managing the state of the shared components of the application and of the information related to the rest of the Meta-Operators that share the same Metaverse. The World Sharing service also manages the anchors, which are entities used in AR/MR applications to align virtual objects in the same physical location (thus, multiple users can see the virtual elements in the same physical position). Finally, the Sharing API deals with the communications with other devices, keeping all the involved parties updated on the events that happen in a shared experience.

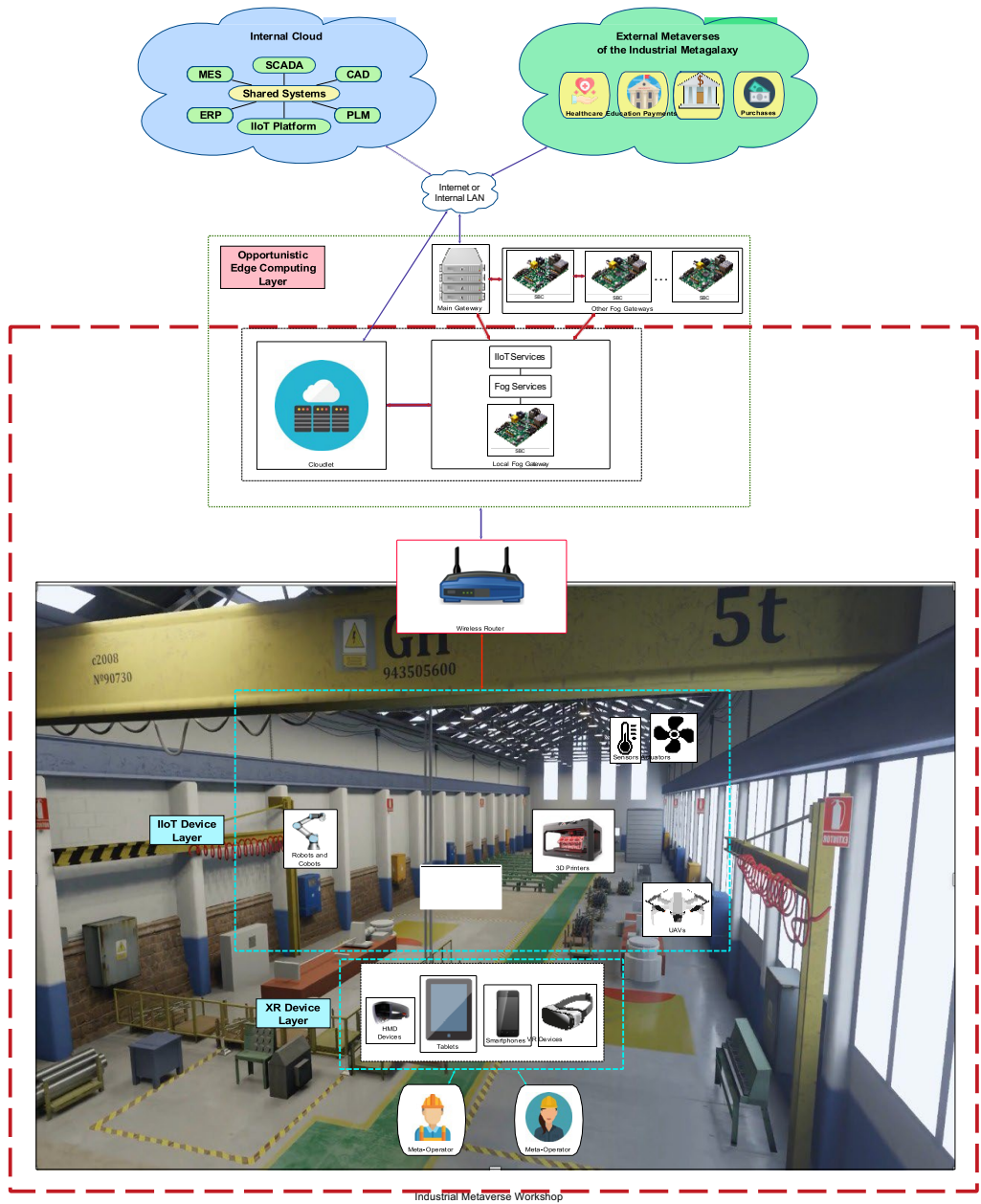

FIGURE 10: Advanced communications architecture for the Industrial Metaverse.

All the XR devices within the same local network connect with each other by using the Sharing API. However, if such devices are not located in the same area (for example, they may be in different rooms of the same building), the communications among them can be performed through the opportunistic Edge Computing subsystem (whose components have been previously described at the end of Section II-D), thus avoiding unnecessary connections to the Cloud, what reduces response latency significantly and, therefore, improves user experience. Nonetheless, in cases where Meta-Operators are far apart and no opportunistic Edge Computing devices are available, the devices can communicate through the Sharing Service available on the Cloud.

Regarding the IoT/IIoT subsystem, it makes use of a service that, depending on IoT/IIoT device location, it can run either on the Cloud or on the opportunistic Edge Computing subsystem. The IoT/IIoT subsystem tracks the state and changes of every IoT/IIoT object, sending notifications to the XR devices when needed, so that they can receive the latest information and events. This subsystem also allows developers to exchange data that feed the XR virtual content, which can react to changes in real-world IoT/IIoT devices. Furthermore, the IoT/IIoT subsystem can send messages back to the real devices in response to the Meta-Operator's interactions (like clicks or when grabbing a specific virtual object).

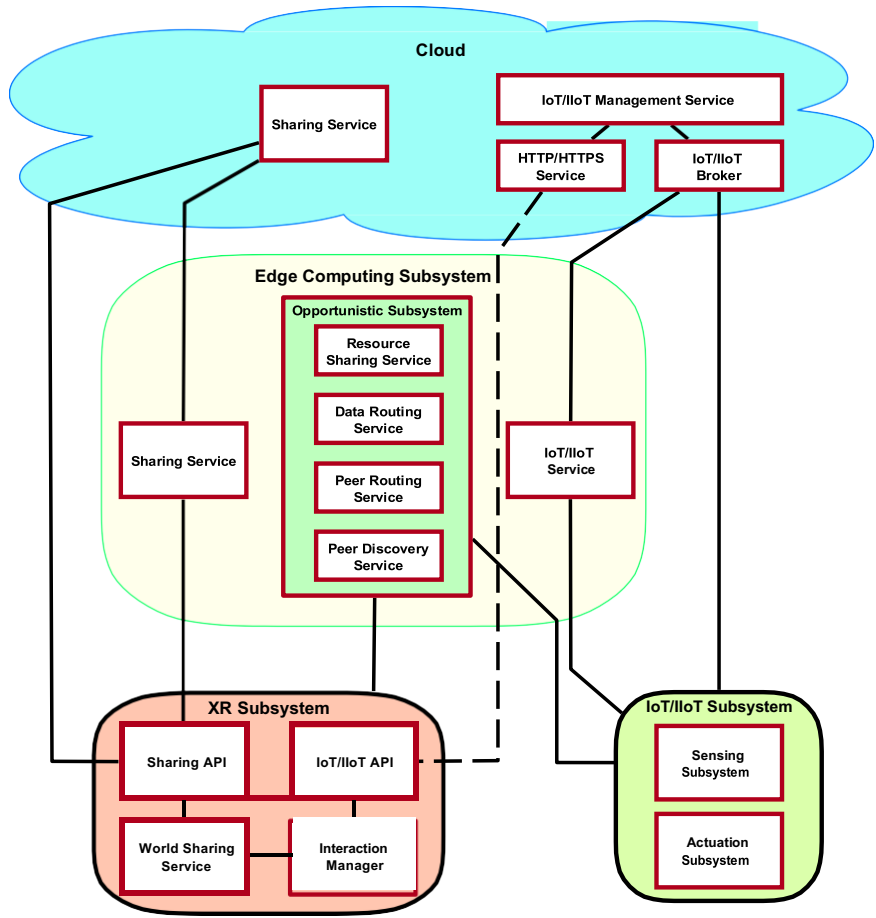

FIGURE 11: Example of the components of a software architecture for implementing Industrial Metaverse applications.

C. SHARED INDUSTRIAL METAVERSE EXPERIENCES

One key point of an immersive Industrial Metaverse is its capability for sharing virtual resources among Meta-Operators in real time. This is not an easy task, especially in industrial environments that change through time dynamically and with which multiple Meta-Operators can interact. In fact, these sharing tasks are so specific that some authors consider it a different subset of XR called 'Shared Reality' (SR) [175].

To share virtual element positions, XR development technologies provide tools to generate a bundle of bytes that represent an anchor. Such a bundle can be stored or shared across multiple devices to establish the same coordinate system throughout all of them. However, to create a shared experience, an application that runs on an XR device has to transfer the bundle over the network and keep track of any changes that each user performs on the environment.

In order to implement an XR sharing service that works without the need of an external coordination server, an opportunistic device discovery service is necessary. Such a service has to detect neighboring devices on the local network and to establish communications with them. For instance, a sharing service can make use of broadcast frames to announce the presence of a new device in the network and to determine the role of each device (Section IV-D provides an example of such a protocol).

Finally, it must be emphasized that the provided shared experiences, in an Industry 5.0 context, should be human-centric, thus delivering an appropriate level of Quality of Experience (QoE), which is conditioned by four main factors [184]: the visual consistency of the experience, its authenticity, its visual smoothness and its comfort.

D. OPPORTUNISTIC COLLABORATIVE PROTOCOL

In order to provide a shared experience, it is necessary to devise collaborative protocols that include opportunistic com-ponents to discover and interact with the surrounding objects. As an example, Figure 12 illustrates how an opportunistic collaborative protocol would operate in a practical Industrial Metaverse scenario. Actually, Figure 12 is a UML sequence diagram whose flow is divided into three stages that represent how three Meta-Operators would interact opportunistically through a collaborative protocol:

- Stage 1: a first Meta-Operator (Meta-Operator 1) would join the network and would start broadcasting HELLO messages to determine whether there are other Meta-Operators in such a network. Since there are none, it assumes the role of network coordinator.
- Stage 2: a second Meta-Operator (Meta-Operator 2) joins the network. The new Meta-Operator sends a HELLO broadcast to the network, which is answered by Meta-Operator 1, who indicates that he/she is the coordinator. As a consequence, Meta-Operator 2 assumes the role of subordinate. Next, it asks Meta-Operator 1 for the shared anchor, which is transmitted as a response.
- Stage 3: a third Meta-Operator (Meta-Operator 3) joins the collaborative experience and Meta-Operators 2 and 3 receive an update from Meta-Operator 1 regarding one of his/her movements in the shared space. In this case, Meta-Operator 3 joins the network like Meta-Operator 2 in Stage 2 (in Figure 12 anchor synchronization for Meta-Operator 3 is omitted just for simplifying the Figure).

Although the previously described opportunistic collaborative protocol may seem simple, it has to deal with different aspects that are essential when developing a collaborative Industrial Metaverse:

- Role assignment. The designed opportunistic collaborative protocol is based on coordinator-subordinate communications where any node can take the role of coordinator or subordinate depending on what the system needs. As a consequence, when a node joins the network, its first task is to determine whether it has to act as coordinator or subordinate. For such a purpose, a HELLO message is sent, which is a broadcast that is only answered by the coordinator device of the network. Thus, if there is a coordinator, the new device acts a subordinate, and if there is not, it acts as coordinator device.
- Initial synchronization. If a new Meta-Operator joins the collaborative network and finds that there are already other Meta-Operators sharing content, after receiving the role of subordinate, it asks the other Meta-Operators for the shared anchor. Since anchors usually consist in one large file (depending on the complexity of the scenario, file size may go from a few to hundreds of megabytes), anchor exchange needs to be carried through transport protocols like TCP to protect and recover automatically from communication errors.
- Synchronization of the interactions. Once all Meta-Operators share the same anchor, their interactions are transmitted to the other Meta-Operators to keep them updated. For instance, such updates can be sent as broadcast UDP packets, since they are usually small (less than 50 bytes) and only convey the most relevant interactions (e.g., user movements, clicks on virtual objects).
- IoT/IIoT integration. As it was briefly mentioned in Section III-E, there are different protocols that can be used to integrate IoT/IIoT devices. For instance, MQTT can be used for exchanging data/requests easily with sensors/actuators [174].

Only a few authors have previously proposed collaborative protocols for AR/MR applications. For example, in [176] it is presented a proof of concept that integrates Microsoft HoloLens with sensor data in a collaborative way. For such a purpose, the authors make use of HoloLens spatial anchors through HoloToolkit, which is currently considered deprecated. Similarly, Microsoft started to develop a sharing framework for collaborative environments [177], but it has not been updated in the last years in favor of Microsoft Mesh [178], which is a platform that allows people in different physical locations to collaborate through diverse devices. In contrast to other collaborative frameworks proposed in the literature [63], Microsoft's solution is focused on off-site mixed collaborative experiences and it relies on Azure, Microsoft's Cloud Computing platform, which involves the disadvantages of current Cloud-based systems (as previously indicated in Section II-D).

Regarding future collaborative AR/MR solutions, it seems that they will be linked to the development of fast wireless communications (e.g., 5G/6G networks) [179] and of the different variants of the Edge Computing paradigm, which will lead to the creation of the Tactile Internet [180], [181]. Although Cloud-based offloading leverages the resources of remote central systems, user experience may be degraded by network delays, while computational cost may be increased when incurring in high concurrent loads [182].

Offloading computing to the edge seems a promising solution [182], [183], mainly as a result of the pervasive deployment of edge servers, which offer additional computing and storage resources for mobile AR/MR applications. Moreover, the execution of local computations can also help to protect privacy and to provide cyber-resilient applications with no single point of failure.

## V. INDUSTRY 5.0 APPLICATIONS FOR THE INDUSTRIAL METAVERSE

The Industrial Metaverse, as a fusion of virtual and physical industrial environments, has the potential to revolutionize diverse aspects of the next generation of factories of the Industry 5.0. Such a potential not only impacts industrial tasks, but also other aspects of the working life of a Meta-Operator (e.g., his/her health or training), which are directly linked to the Industry 5.0 foundations. As an illustration of the potential of the Industrial Metaverse to revolutionize the Industry 5.0 factory, the following subsections describe examples of human-centric applications for Meta-Operators.

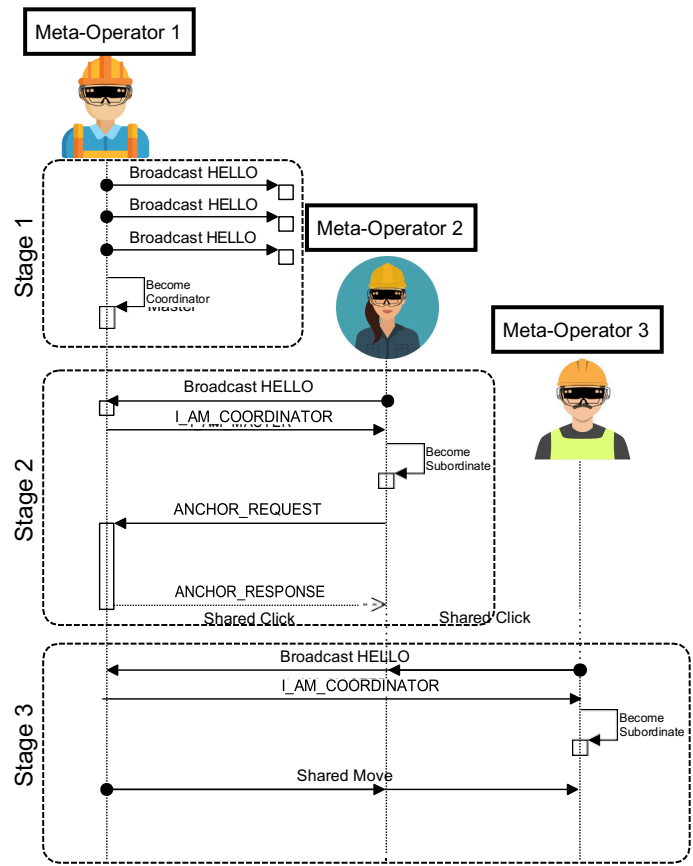

FIGURE 12: Example of flow diagram of the execution of an opportunistic collaborative protocol for an Industrial Metaverse.

### A. GUIDED MANUFACTURING

In areas like the automotive industry or heavy industries, manufacturing is a complex process that consists of many stages that require to perform really precise steps to meet the expected product quality. AR and MR devices help to follow step-by-step assembly instructions with a precision that can be superior to the one obtained through traditional manufacturing processes [185]. Thus, Meta-Operators can join Industry 5.0 manufacturing metaverses where they are guided through overlaid 3D models, animations and videos along the different manufacturing stages [186]. Moreover, Meta-Operator's UX can be significantly enhanced by providing attractive virtual user interfaces that do not require to make use of a computer or of input devices like a mouse or a keyboard, which are frequently difficult to use when carrying heavy weights or when wearing gloves.

As an example, Figure 13 shows two Meta-Operators that execute an application in a shipyard that belongs to Navantia (the largest shipbuilding company in Spain and one of the ten largest shipbuilders in the world [187]), which tested the developed system in its turbine workshop. Specifically, the application is described in [64] and allows for sharing virtual content for training and guiding Meta-Operators during the assembly of a hydraulic clutch of a turbine.

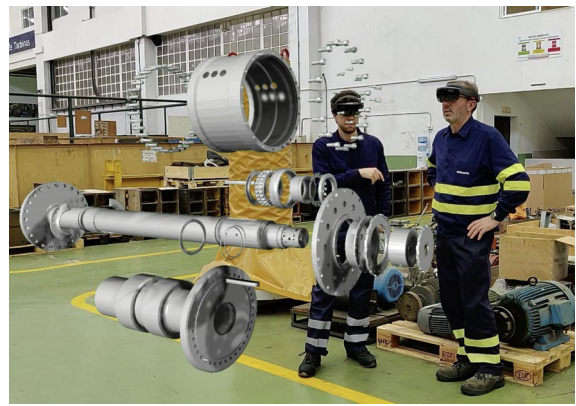

FIGURE 13: Meta-Operators using a guided manufacturing application developed for Navantia's Metaverse.

### B. OBJECT IDENTIFICATION AND TRACKING

To interact with the physical world in an advanced Industrial Metaverse, it is necessary to somehow identify the surrounding objects and then track them. In this regard, it is possible to develop Industrial Metaverse applications that make use of item, vehicle and tool tracking mechanisms through AI-enabled object detection techniques [188] and with the help of industrial Auto-Identification [189] or Real-Time Location Systems (RTLSs) [190]. In fact, RTLS systems enable using technologies like RFID in industrial environments to identify the surrounding items [75]. In addition, the same object identification/tracking systems usually allow for visualizing certain information on the state of the monitored objects or of their sensors/actuators.

Other authors proposed to identify and track certain visual cues to provide guidance. For instance, in [191] it is proposed to detect parking spots or lanes to provide drivers assistance in real time through a Metaverse application.

### C. SUPPLY CHAIN AND LOGISTICS

Despite the challenges to involve supply chain entities in adopting the Industrial Metaverse technologies [192], industrial companies and their logistics can be benefited by the use of Industrial Metaverse applications, since they have the potential to enhance the agility and adaptability of supply chain operations for sustainable business models [193]. Moreover, payment and other banking operations can be easily translated to the Industrial Metaverse thanks to the growing use of digital currencies [24], [54].

An example of Industrial Metaverse based supply chain has been already proposed by Renault, who is looking for using a digital twin of each car part so that suppliers can produce physical copies [165]. Other authors have also proposed similar approaches for the Japanese automotive industry [194]. Such previous solutions can be easily adapted to develop Industrial Metaverse applications based on digital twins of warehouses in order to facilitate their management, which requires to identify, to locate and to sort the stored items. In fact, having an efficient and agile warehouse management process is essential for creating Industry 5.0 solutions [195], which can be also benefited by the object identification/tracking technologies mentioned in the previous subsection.

### D. QUALITY CONTROL

Quality control is essential for Industry 5.0 factories that build complex products. Nowadays, most quality control procedures are carried out manually by industrial operators through visual inspections that verify whether the products pass the established quality requirements.

The concept of Industrial Metaverse can help such quality controls by automating part of them through the use of advanced AR/MR devices to inspect the products [197]. For instance, in [198] it is detailed an ML-based system that is able to classify as 'defective' or 'non defective' the items built by a central processing unit system production line. Another image-based solution that can be applied in an Industrial Metaverse is presented in [199], where the authors describe an AI-based solution that makes use of Deep Learning (DL) and a custom Convolutional Neural Network (CNN) to automate the visual inspection of casting products.

Specifically, the proposed system achieves an accuracy of 99.86% in the evaluation industrial environment selected by the authors.

### E. MAINTENANCE

Maintenance consists of a set of processes that combine technical, administrative and managerial actions during the life-cycle of an item in order to preserve its state and required functionality. The Industrial Metaverse can be really useful by automating part of such tasks (e.g., through AI-enabled interconnected XR-based applications) and enhancing them by providing real-time data and 3D content, which goes farther than traditional paperwork. For instance, in [200] it is described the application of the Industrial Metaverse to railway maintenance through the use of XR headsets (HTC Vive, Oculus Quest 2, Microsoft HoloLens 2), a smartphone-based LiDAR and AI techniques. Similarly, in [201] it is presented an AR/MR-based system for carrying out the maintenance of machine tools.

### F. IOT/IIOT DEVICE INTERACTION

As it was previously indicated in Section III-E, AR and MR can interact with IoT/IIoT devices. Such an interaction is performed by AR/MR devices by sending requests to the deployed IoT/IIoT machines and then by processing the responses received from them. However, there is a difference between AR and MR: while AR simply collects and shows data related to the IoT/IIoT devices, MR is also able to interact with them, thus impacting the physical world where Meta-Operators work.

IoT/IIoT interaction is usually carried out through virtual panels and is often linked to the digital twins of the IoT/IIoT devices. For instance, in [202] it is described the digital twin of a smart lamp. Such a digital twin was created thanks to the use of an IoT-enabled smart power outlet and is able to switch on and off the plugged appliances and measure their power consumption and environmental temperature. In the developed system, the IoT subsystem updates the information on the smart lamp in real-time and shows it through a virtual panel. In addition, users can interact with the virtual model, thus sending commands to the real physical object. Such features are illustrated in Figure 14, which shows the main menu of the developed HoloLens application. On the left, a set of buttons allows for controlling the different actions that can be performed. In the middle, the real-time data collected from the power consumption and temperature sensors are shown. On the right of Figure 14 is the remotely operated smart lamp.

Robotics is also an area were the Industrial Metaverse can be useful, since it involves integrating physical devices (i.e., sensors and actuators) and digital remote control interfaces, which jointly has been coined by some authors as 'Metarobotics' [203]. For example, in [204] it is described a framework to minimize the required packet rate when controlling a robotic arm in the Industrial Metaverse. The control of robotic arm is also described in [205] together with other three use cases that fuse robotics, XR technologies and digital twins.

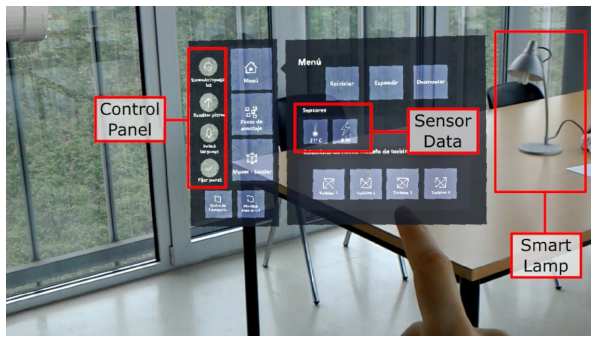

FIGURE 14: Meta-Operator controlled IoT smart lamp.

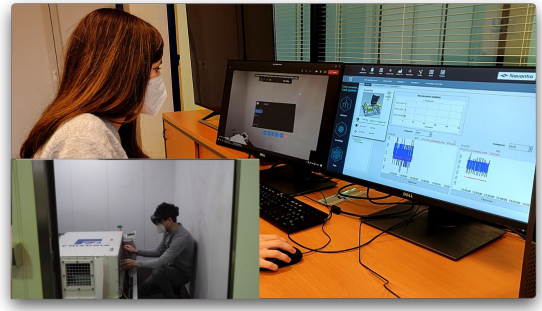

FIGURE 15: Example of remote assistance to a Meta-Operator.

### G. REMOTE ASSISTANCE

The technologies behind the Industrial Metaverse make possible to provide remote augmented communications with other Meta-Operators or with users that make use of computers or mobile devices. Such a kind of communications is useful in large industrial facilities where fast responses are necessary. Although smartphones can provide voice, text, pictures and video communications, such exchanges are prone to misunderstandings and to human errors (i.e., they depend on manual tasks carried out by the operators).

The Industrial Metaverse can immerse the communicating parties into augmented personal communications, which can overlay virtual elements to reality to deliver accurate indications. In addition, AR/MR devices embed cameras that allow Meta-Operators to share their point of view to show what they are seeing and the elements that exist in the operators' surroundings. Furthermore, many Industrial Metaverse devices can record the audio and video exchange during the remote assistance call to document in a transparent way how the issue that originated the call was handled.

As an example, Figure 15 illustrates the use of an application that allows for communicating a Meta-Operator with a remote support worker. The Figure shows what the female operator watching on a PC live video sent by the remote Meta-Operator through his Microsoft HoloLens smart glasses. At the bottom of the Figure, on the left, it is shown such a remote Meta-Operator, who can use his voice like in a regular call, but also send pictures and include visual indications to the female operator that is providing assistance.

Another example is described in [206], where the authors make use of Microsoft HoloLens 2 and Meta Quest 2 to enable remote experts to assist local technicians through a MR/VR joint Industrial Metaverse. Thus, both MR and VR users have their own avatars and can interact with each other through audio (they use bone conduction headsets) and video (VR experts can receive a live video stream from the local technician).

### H. REMOTE MAINTENANCE

Remote assistance can be applied to the specific case of performing remote maintenance tasks. It is important to note that most machinery and industrial tools require to be maintained and calibrated periodically. AR and MR devices can provide Industrial Metaverse applications to help to carry out maintenance procedures either physically (when human presence is necessary) or virtually (when maintenance only requires remote software verifications and updates). In the former case, AR/MR-based applications can show virtual interfaces that allows for guiding Meta-Operators through step-by-step instructions (e.g., to indicate how to disassemble the maintained machine and how to perform with precision the maintenance procedure). As an example, Figure 16 shows an AR application for Microsoft HoloLens developed by the authors of this article for the remote and collaborative maintenance of a hydraulic clutch. Another example of remote assistance is described in [207]. In such a work an operator wears a camera on his helmet to transmit live video to a remote assistant that watches it through a VR headset.

### I. HIDDEN AREA VISUALIZATION

In some industries it is necessary to 'see through structures' to visualize their inner components. For instance, in the construction [208] or in the shipbuilding industry [12] it is necessary to be able to see the piping or wiring that is behind walls or the ceiling before performing certain maintenance or repair tasks. In such cases, Meta-Operators can make use of AR/MR applications to see behind the structure by visualizing the position of the piping/wiring as virtual elements before proceeding with the disassembly of the structure.

### J. ENHANCED INTERACTION WITH INDUSTRIAL SOFTWARE

The Industry 4.0 paradigm supposed the digitalization of many traditional industries, which updated their software to advanced solutions based on, for instance, PLM, ERP or MES software. Such software is often designed and executed on regular computers, smartphones or tablets, so its UX is optimized for such a kind of devices. However, such software can be ported to be used as Industrial Metaverse applications, making their interface more attractive for Meta-Operators that would use it through AR/MR devices.

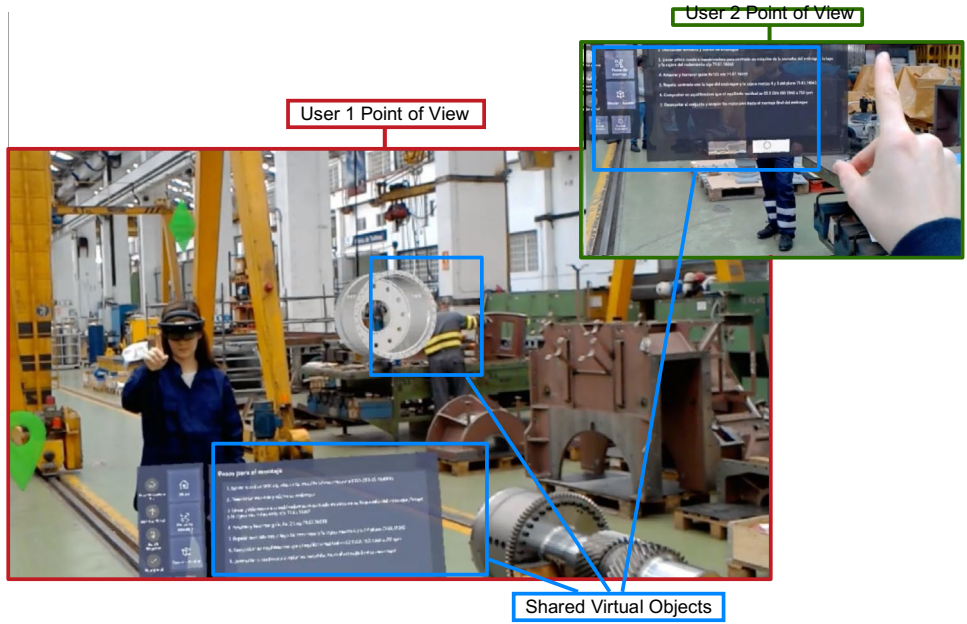

FIGURE 16: AR/MR application developed for the maintenance of a hydraulic clutch.

As an example, Fig. 17 shows an application for Microsoft HoloLens that is able to show ship blueprints both in 2D and 3D (at a real scale). Another practical example is described in [26], where the authors describe a VR-based application that provides virtual dashboards fed with real-time data from a remote Cloud to manage a sewage sludge plant.

### K. COLLABORATIVE DESIGN

Many industries perform design tasks in a collaborative way (e.g., in the automotive or construction industries). The Industrial Metaverse provides an excellent platform for carrying out such tasks For example, Figure 18 shows an example of an application for Meta-Operators that enables visualizing and interacting collaboratively with the structures to be placed on the land where a factory will be built.

An example of collaborative product design application is described in [209]. Specifically, the paper describes how the authors created a VR-based Industrial Metaverse where engineers and clients can meet virtually to conduct interactive product design reviews.

### L. ENHANCED AND COLLABORATIVE TRAINING

Education is one of the fields that can be more benefited from the progress made on Metaverse technologies [23], [210]. In the specific case of the Industrial Metaverse, realistic experiences can be created for training future Meta-Operators, thus avoiding the need for dedicating experienced operator time to the trainees [211]. Moreover, in many cases, when the oldest operators retire, their skills and experience, which are often not properly documented, are lost.

To address such an issue, it is possible to develop AR/MR applications based on the input from skilled operators to train future Meta-Operators in attractive augmented, mixed or entirely virtual learning environments. Moreover, such applications can monitor and quantify the progress of the trainees and thus evaluate their learning outcomes.

The developed Industrial Metaverse can also be really useful when the resources to train future Meta-Operators are expensive. For instance, AR/MR/VR applications that provide simulations, from flight training to combat scenarios, offer cost-effective alternatives to real-world experiences [212], [213]. Specifically, in military training, AR/MR environments recreate diverse conditions, providing valuable hands-on experience [214], [215].

### M. REMOTE HEALTH ASSISTANCE

Health is essential for fulfilling the human-centric nature of Industry 5.0. The Industrial Metaverse can suppose an attractive and agile platform for providing healthcare consultations [166]. Potential patients can access services globally, and doctors may utilize smart IoT/IIoT devices for diagnostics [216]. Moreover, virtual reality healthcare aides to provide real-time feedback on daily work activities [217].

Furthermore, in certain cases where industrial operators may face physical or mental illnesses, the Industrial Metaverse technologies can be adapted to consider such conditions. In fact, AR and VR have been previously used in therapeutic applications, particularly in the treatment of post-traumatic stress disorder (PTSD) [218] or in rehabilitation [219].

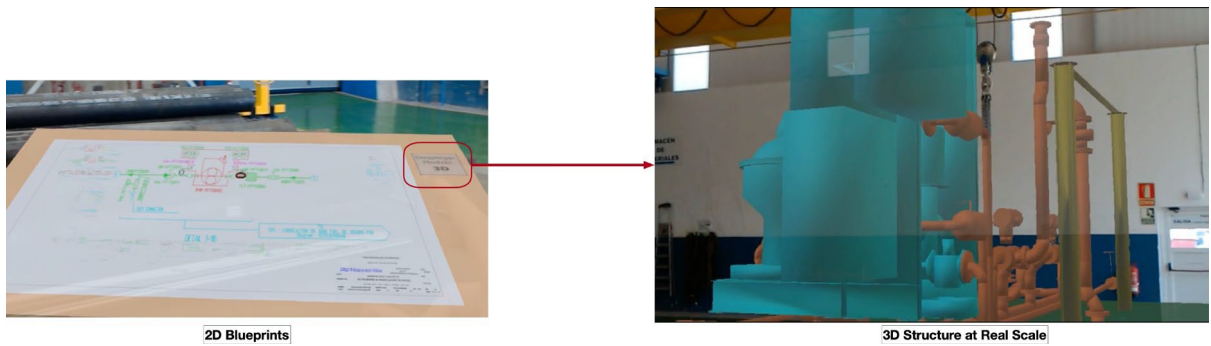

FIGURE 17: Meta-Operator application for visualizing ship blueprints.

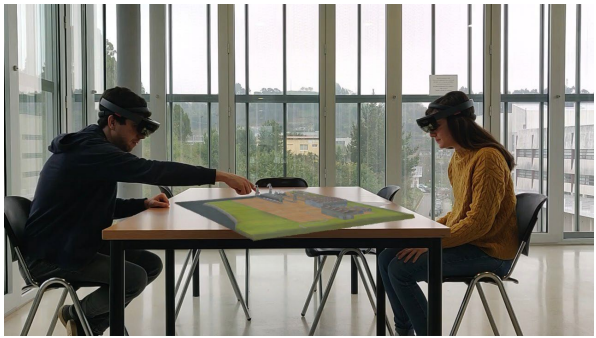

FIGURE 18: Collaborative design between Meta-Operators.

### N. ADVERTISING AND MARKETING

Advertising evolves into immersive experiences in the Industrial Metaverse. Advertisers leverage detailed user data to personalize advertisements, offering a more engaging and relevant experience to consumers. For instance, industrial companies can provide Industrial Metaverse applications for [220]:

- VR exhibition halls to show their products to remote customers.
- AR product catalogs that make use of smartphones, tablets or AR glasses to interact with a virtual representation of the products.
- MR layouts of the products that allow for placing them in real locations in order to determine whether they are an appropriate fit for the environment and operation conditions.

### O. OTHER ENTERPRISE ACTIVITIES

The implementation of Industrial Metaverses will involve significant organizational changes in companies [221] and to deal with new challenges related to corporate responsibility [222]. For instance, industrial companies are currently facing talent scarcity in many places around the world, being difficult to attract and hire the best engineers and opera- tors.

In such a scenario, the Industrial Metaverse can help recruitment by providing virtual spaces to attract talented individuals and hold meetings, creating virtual communities.

Companies can also explore virtual worlds for product design, idea testing, and collaborative innovation. Virtual workplaces, mixed reality interactions and online education in the workplace will become commonplace [54]. In fact, some authors already use the concept of 'Metaverse office' [223] and desktop-simulated testing environments to perform remote assembly tasks [224].

Moreover, it must be mentioned that many companies have established museums to transfer to the general public their history and achievements. In such situations, museums can embrace XR to provide immersive historical experiences through personalized metaverses or by being part of generic platforms [140]. XR technologies enable users to witness and interact with historical events, fostering a deeper understanding of various eras. Such aspects can be benefited by the story telling capabilities provided by XR technologies, which provide more immersive experiences than traditional 2D-content based museums [225].

## VI. CHALLENGES
### A. MAIN DEVELOPMENT CHALLENGES
#### 1) Hardware cost

Although some studies indicate that Industrial Metaverse technologies will not be mature until around 2030 [39], the current price of the latest XR devices is still too high for their massive adoption. As it was previously indicated in Table 3, most XR devices that can be used in industrial scenarios usually worth more than $1,000. In this regard, the progress made in XR technologies and the adoption of such technologies by the general public can help to reduce price through scale economies.

#### 2) Need for faster communications and higher bandwidth

The development of the Metaverse necessitates advancements in communication technologies to ensure seamless interactions and a high-quality UX, which is conditioned by fast reaction times. For instance, VR devices require a transmission rate beyond 250 Mbit/s and a data error rate between $10^{-1}$ and $10^{-3}$ to prevent communications interruptions, while haptic devices need a data transmission rate of 1 Mbit/s and a maximum latency of 1 ms [226].

Thus, future ultra-fast reaction times are expected to enhance dramatically human-to-machine interaction, since they will enable building real-time interactive systems for fields like industrial automation, healthcare or gaming [227]. This is especially important in XR development, since they rely on head and eye movements, which have stringent latency requirements to provide a realistic and comfortable UX.

In interactive experiences, low latency is crucial to create a responsive and immersive environment. The human threshold for latency is low, especially in activities like gaming, where delayed responses can impact user experience [228]. Latency, often compounded by jitter (i.e., the variance in delivery time), poses challenges in interactive experiences, leading to delayed responses and potential disruptions in fast-paced activities. Industrial Metaverse developers can take note on the progress made by online gaming to reduce latency, like the use of regional servers or netcode solutions, which include delay-based and rollback netcodes that help synchronize player interactions and maintain consistency in gameplay.

Latency can be identified as a significant networking obstacle on the path to the implementation of Industrial Metaverses, but the lack of widespread demand for ultra-low-latency services today makes it challenging for network operators and technology companies to focus on real-time delivery. As Industrial Metaverses grow, there is an expectation for increased investment in lower-latency Internet infrastructure to support the evolving communication needs.

5G networks, with their promise of ultra-low latency, are positioned to address latency challenges, offering potential improvements compared to 4G networks. Thus, the evolution of technologies like 5G&Beyond technologies is crucial for facilitating processes within the Industrial Metaverse, especially in bridging the gap between in-person and online interactions [206]. In fact, some authors have already proposed to design 6G communications to enable building robust digital twins, which will based on decentralized network architectures that use Machine Learning to deliver Internet of Everything (IoE) services [229] (other authors also include besides 6G the use of Edge-AI as another key foundation for the future Industrial Metaverse [72]). Similarly, in [230] the authors analyze the fundamental supporting role of 6G in order to create future Industrial Metaverses.

For remote physical locations, satellite-based solutions like Starlink [231] aim to provide high-bandwidth, low-latency Internet across the globe, but challenges such as latency increase over long distances remain.

Finally, it is worth pointing out that new communication paradigms will need to be explored, like task-oriented communications [232], which can help to optimize wireless communications when executing typical tasks of an Industrial Metaverse like real-world virtualization, virtual-world projections and cooperative-AI training.

### 3) Cybsersecurity challenges

The emergence and growth of the Industrial Metaverse present both new vulnerabilities and opportunities for cyber-attacks. As the Metaverse progresses in the integration of advanced industrial technologies, it will become imperative to address cybersecurity concerns and protect Meta-Operator data [90]. Moreover, it is important to note that the Industrial Metaverse involves protecting both physical and virtual assets whose interactions may be really complex, so it is necessary to contemplate new cybersecurity paradigms that protect both worlds in parallel, feeding the inputs from one world into the other, instead of considering them independently [233].

There are also other key aspects to be considered to guarantee cybersecurity in the Industrial Metaverse:

- Dark patterns. Dark patterns refer to deceptive user interfaces designed to manipulate users into making choices that may not be in their best interest [234]. In the Industrial Metaverse, the potential for dark patterns arises as immersive experiences could be crafted to exploit users' perceptions and behaviors [235].
- Biometric information. XR devices incorporate sophisticated cameras, face recognition, eye tracking, body tracking and even electroencephalogram (EEG) technology to detect brainwave patterns [236], [237]. This extensive data collection poses privacy risks, and the enormous amount of data generated becomes a valuable target for cyberattacks [238]. Protecting this sensitive information from unauthorized access and misuse is a critical cybersecurity challenge.
- Post-Quantum security. The advent of quantum computing poses a potential threat to existing cryptographic methods [239]. Post-quantum security measures will be essential to ensure the resilience of Industrial Metaverse platforms against advanced cyber threats [240]. Therefore, updating cryptographic protocols and adopting quantum-resistant algorithms become imperative components of cybersecurity strategies.
- Decentralized/Distributed architectures. The question of whether the Industrial Metaverse operates on a decentralized or distributed basis raises challenges related to user permissions [171]. In decentralized models, obtaining and providing explicit user consent for data usage becomes complex. Ensuring that cybersecurity measures are uniformly applied across decentralized networks is essential for protecting user privacy and maintaining ethical data practices. Moreover, IoT/IIoT integration into the Metaverse is a challenge that can be tackled through decentralized/distributed approaches [241].
- Government regulations and surveillance. The example of China's regulatory measures, such as restricting minors' gaming hours and using facial recognition for verification, highlights the intersection of government regulations and cybersecurity in the Metaverse [242]. Thus, balancing regulatory compliance with user privacy and protection from potential surveillance practices becomes a critical aspect of cybersecurity in the Industrial Metaverse.

4) Interoperability

Achieving a unified Industrial Metaverse is currently difficult, existing multiple competing networks of virtual worlds that require to be connected [243]. This echoes past debates during the 'Protocol Wars', in the early days of the Internet, when there was uncertainty about establishing a common internet standard [244]. Overcoming these debates and achieving consensus on interoperability standards is crucial.

The following are some of the most relevant key considerations and challenges related to achieving interoperability:

- Data exchange and permissions. Data gathered in the Industrial Metaverse must move easily across different platforms and operators to enable interoperability. Establishing bilateral and multilateral permissioned agreements between software developers and enterprises becomes necessary. However, ensuring secure data exchange while respecting privacy and user permissions adds complexity.
- Definition and standards. Virtual worlds need to be interoperable, allowing systems and software to exchange information seamlessly. The challenge lies in defining common standards for data exchange within the Industrial Metaverse. Unlike the Internet, where standards like TCP/IP enable global information exchange, the existing virtual worlds still lack a common language for communication.
- Isolation and fragmentation. Most popular virtual worlds today use their own rendering engines, file formats, and unique systems, leading to isolation and fragmentation. Lack of interoperability results from virtual worlds being designed as closed experiences with controlled economies, optimized for specific purposes rather than collaboration.
- Complexity of avatars and objects. The complexity of defining interoperable avatars and objects in a 3D space adds a layer of difficulty. Questions arise about the structure of avatars, including clothing, accessories, and movement characteristics. Developers need to agree on standards that define the components of avatars and objects in a coherent and comprehensive manner.
- User expectations and diversity. Users have diverse expectations regarding avatars and objects in the Industrial Metaverse. From anthropomorphic avatars to inanimate objects, different categories have unique characteristics and behaviors. Developers must understand and agree on how avatars and objects should function, move, and interact within the virtual environment.
- Ownership. Handling ownership records of virtual goods across multiple virtual worlds raises challenges. If a company purchases a virtual good in one world and uses it in others, managing ownership records and updating them becomes complex.
- Handling virtual goods. Sharing and validating ownership records of virtual goods across diverse virtual worlds require standardized protocols. Developers need code that can interpret, modify, and approve third-party virtual goods. Monetization models, validation processes, and ownership management need to be defined for virtual goods that move across Industrial Metaverses.

5) Need for 3D content and rendering

In the Industrial Metaverse, the use of 3D content is essential, since it serves as a catalyst for moving from the physical realm to a truly immersive digital domain. For such a purpose, several aspects need to be considered:

- The evolution beyond 2D. The essence of the Industrial Metaverse lies in its departure from the familiar 2D constructs of the current Internet. Message boards, chat services and interconnected networks have long been part of the daily routing of industrial operators. However, the development of 3D environments is imperative for the metamorphosis of human interaction in digital spaces [245]. In fact, some authors emphasize the intuitiveness of 3D as an interaction model [246], especially in social contexts, in some cases arguing that humans did not evolve for millennia to engage with flat touchscreens.
- Digital evolution. Human inclination leans towards digital models that closely mirror real-world richly detailed experiences (e.g., with high quality audio and video) and delivering a sense of being 'live'. As online experiences become more immersive, real lives will migrate to the digital realm, shaping also corporate culture.
- Towards the '3D Internet'. If the trajectory towards a '3D Internet' unfolds, it holds the potential to disrupt traditionally resistant industries. For instance, industries like Education might find its catalyst in the enhanced capabilities of 3D virtual worlds and simulations [23]. The traditional barriers to remote education may crumble, ushering in a new era where students worldwide can immerse themselves in virtual classrooms, partake in interactive experiences and revolutionize pedagogical practices.
- Coexistance of 2D and 3D content. While the Industrial Metaverse is envisioned as a 3D experience, it does not mandate that everything within it adheres to a 3D format. The coexistence of traditional 2D interfaces alongside 3D experiences is anticipated. Therefore, the transition to a 3D Industrial Metaverse does not imply a universal shift for the entirety of the Internet and computing.
- Immersive VR as an accessory, not a necessity. It is crucial to dispel the notion that immersive VR headsets are a prerequisite for participating in the Industrial Metaverse. While VR may become a popular mode of access, it is just one facet of the multifaceted entry into the Industrial Metaverse.

6) Development of Industrial Metaverse-ready IIoT devices

The evolution of the Metaverse is intricately tied to the evolution and proliferation of IIoT and IoT devices [247].

Such devices, when embedded appropriately in an industrial environment, hold the potential to transform various industries, offering a seamless connection between the physical and virtual realms [163]. Nonetheless, the development and deployment of IIoT devices need to consider several factors for creating Industrial Metaverse-ready solutions:

- IoT/IIoT device integration with XR systems. It is currently not straightforward to integrate IoT/IIoT devices with the visualization devices required to access the Industrial Metaverse since communication protocols are usually different, so adaptations need to be carried out. Ideally, IoT/IIoT devices should incorporate plug-and-play protocols like the ones associated with Transducer Electronic Datasheets (TEDs) [248].
- IoT/IIoT device interoperability. IoT/IIoT devices need to be flexible enough to support different protocols in order to be able to adapt to the requirements of the implemented Industrial Metaverse. In this regard, standardization is essential.
- IoT/IIoT device efficiency in Industrial Metaverse applications. Many IoT/IIoT devices make use of batteries or rely on constraint power sources (e.g., energy harvesting, non-continuous renewable energy sources), so they are not necessarily available the whole time for data exchanges. Therefore, to optimize energy consumption, the use of Green IoT/IIoT strategies is highly advisable [76].
- IoT/IIoT-XR opportunistic communications. There is a clear lack of opportunistic solutions to integrate in real time IoT/IIoT devices with other devices worn by Meta-Operators [20]. Further research is necessary in order to create adapted protocols for IoT/IIoT device discovery, opportunistic mobile data and peer routing, and for providing resource sharing services.
- Fast processing for large amounts of data collected from IoT/IIoT devices. Although the evolution of IoT/IIoT technologies and the existence of new advanced communications architectures (e.g., Edge Computing) have allowed for off-loading part of the computing tasks that have been traditionally carried out locally or in the cloud, more research is needed to process large amounts of data while preserving the latency restrictions of the Industrial Metaverse. Thus, infrastructure deployed locally in factories and workshops like Cloudlets, which may include powerful Graphics Processing Units (GPUs), can be really helpful [69],[71].

### B. MAIN EFFICIENCY CHALLENGES
#### 1) Real-Time Rendering
Real-time rendering is essential for guaranteeing the responsiveness of virtual worlds. Such a process involves solving complex equations and processing inputs and data with the help of computing resources like GPUs and central processing units (CPUs). Processing is usually carried out at a minimum of 30 frames per second (being ideally 120 frames per second), which can become really intensive when dealing with demanding 3D content like large factories or detailed industrial machinery. Thus, in the last years, different techniques have been applied to improve real-time rendering, like polygon-reduction algorithms [249] or triangle mesh reconstruction [250]. In addition, more techniques need to be explored and tested, such as eye-tracking based foveated rendering [251], neural supersampling [252] or advanced virtual view synthesis techniques that use a small number of input views [253].

It is important to note that real-time rendering and Industrial Metaverse scalability are dependent, so the previously mentioned techniques will also help to provide an adequate experience to a growing number of Meta-Operators that can make use of multiple virtual worlds.

#### 2) Persistence of large amounts of data
The virtual worlds of the Industrial Metaverse require data retention, rendering and sharing to deliver a coherent experience where virtual elements preserve their location and state. The challenge lies in balancing the desire for detailed persistence against the computational demands and resource constraints. In fact, the sheer volume of data involved in managing with high fidelity the persistence within the Industrial Metaverse surpasses current technological feasibility. In fact, this challenge is so important that the IEEE has already established a specific working group to study the different technologies that will be required to tackle this issue [103]. For instance, such a need for persisting high volumes of data requires to explore solutions like graph-based NoSQL databases, which overcome some of the limitations of relational databases [254]. Moreover, the application of Big Data techniques should still be analyzed carefully for the specific case of the Industrial Metaverse [255].

#### 3) Synchronization
In the dynamic landscape of the Industrial Metaverse, synchronization emerges as a pivotal challenge, dictating the feasibility of shared virtual experiences. Synchronization requires high-bandwidth, low-latency and continuous Internet connectivity, which jointly suppose a challenge for seamless and immersive experiences:

- High bandwidth. To enable the transmission of substantial data volumes within a designated timeframe, participants in a virtual world must possess high-bandwidth connectivity. This component ensures the fluid exchange of information, underpinning the richness of shared experiences.
- Low latency. The demand for a low-latency connection, synonymous with swift responsiveness, becomes imperative. In the realm of synchronous online experiences, delays are intolerable; user inputs must seamlessly translate into real-time reactions within the virtual world.
- Continuous connectivity. Synchronization mandates a sustained and uninterrupted connection between every

participant and the virtual world. The continuity of this connection, both in the transmission and reception of data, is non-negotiable. Interruptions or lapses could shatter the cohesiveness of the shared experience.
- The Internet's design constraint. The foundational challenge underlying synchronization lies in the fact that the Internet, at its core, was not crafted for synchronous shared experiences. Instead, its origins trace back to a framework designed for the dissemination of static copies of messages and files that were accessed by one party at a time, primarily within research labs and universities.

4) Better visualization hardware

The quest for immersive experiences through XR devices confronts a crucial bottleneck with the visualization hardware, which is years away from expectations [256]. This challenge unfolds across dimensions of resolution, refresh rates, field of view and the intricate interplay between hardware and user experience:
- Resolution and refresh rates. Hardware has progressed significantly in the last years in terms of resolution. For instance, Oculus Quest 2 reached a resolution of 4K per eye, but some authors advocate for resolutions exceeding twice 4K and for the application of additional processing techniques to avoid pixelation problems [257]. Regarding refresh rates, they often oscillate between 72 Hz and 120 Hz. Motion sickness, experienced by a notable percentage of the users, emphasizes the need for higher refresh rates, with 120 Hz posited as the threshold to prevent disorientation.
- Weight. Visualization devices should be as light as possible in order to provide comfort to the Meta-Operators that need to wear them for hours.
- Battery life. Industrial Metaverse device battery life should be at least equal to the number of hours required in a work shift, which may vary from one industry to another, but that should be around 8 hours. Nonetheless, it is important to consider the fact that Meta-Operators do not need to be using continuously an XR device (i.e., they can use it only for performing specific tasks), so practical battery life can be lower than a work shift.
- Field of view. This parameter impacts UX, so the higher the FoV, the better the UX.
- Processing power. More powerful hardware will help to avoid asking for computational resources to external Industrial Metaverse infrastructure (e.g., to the deployed OEC devices or to the Cloud) and will lower response latency, thus improving the Meta-Operator UX. However, it is important to note that more powerful hardware usually consumes more energy, so Industrial Metaverse developers should design systems with an appropriate balance between computing power and battery life. Such a recommendation can be extrapolated to the creation of sustainable Industrial Metaverse solutions that comply with the Industry 5.0 foundations [258].

5) Better interaction devices

The evolution of the Industrial Metaverse relies not only on immersive headsets but also on a plethora of complementary hardware, presenting a multifaceted challenge in designing interaction devices. As it was previously described in Section III-C, XR accessories extend from haptic gloves to bodysuits, but more futuristic concepts are still being developed to push the boundaries of user engagement when interacting with virtual content:
- Smart contact lenses. Concepts like smart contact lenses open new frontiers for deploying XR technology, avoiding the need for carrying cumbersome helmets or unnecessary glasses [259]. Moreover, smart contact lenses, remain a topic of interest due to their potential to measure certain health parameters, like blood glucose concentration [260].
- Brain-computer interfaces (BCIs). BCIs, exemplified by Neuralink [261], venture into the realm of direct brain implants. However, BCI development faces both technical and ethical issues. Moreover, ethical considerations surrounding thought-reading devices, permanence and user consent pose challenges in the market adoption of BCIs.
- Ultrasonic-based haptic interfaces. They emit sound waves, forming a 'force field' in the air, enabling users to sense and interact with virtual entities.
- Advanced motion capture and gesture recognition devices. Gloves, bodysuits, and tracking cameras capture users' motion data, allowing for the real-time reproduction of body movements in virtual environments.
- Advanced wearables. For instance, CTRL-labs, which was recently acquired by Facebook, pioneered electromyography-based armbands that record muscle activity. These devices translate intricate gestures, such as finger movements, into virtual interactions, showcasing the potential for detailed control.

C. OTHER CHALLENGES

1) Standardization

The journey of standardization within the Industrial Metaverse draws parallels with the foundational development of today's Internet. This intricate process will probably take years and involves collaborative efforts from government research labs, universities and independent technologists and institutions.

The following are some of the critical aspects that will impact the dynamics of the standardization of the Industrial Metaverse:
- Evolution requires collaboration. The creation of the Industrial Metaverse should harness the lessons learned during the evolution of the Internet, which was nurtured through collaborative endeavors, featuring consortiums and informal working groups. Comprising not-for-profit entities, these collectives shared a common goal: to craft open standards fostering collaboration on diverse tech-nologies, projects and ideas. The inclusivity of this approach allowed for a diverse range of voices, from government labs to independent technologists, contributing to the Internet evolution.

- Democratizing Industrial Metaverse creation. The widespread adoption of common standards will proliferate in a democratized era where anyone with an Internet connection can create Metaverse content swiftly and at minimal cost. The simplicity of using Industrial Metaverse platforms will empower users to create content accessible across XR devices and users worldwide. As it occurred with the Internet, standard universality will eliminate barriers, enabling seamless communication and collaboration.
- Economic and collaborative benefits. Standardization can reap multifaceted benefits, extending beyond technical considerations. The economic advantages are evident as it can become more cost-effective and straightforward to engage with external vendors, integrate third-party software and repurpose code. The open-source nature of many standards should foster an ecosystem where individual innovations reverberate throughout, simultaneously challenging proprietary standards and mitigating the dominance of intermediary platforms.
- Empowering company-driven content. Common standards can ensure that Meta-Operators and developers remain at the forefront, eliminating the need for disintermediation. This can empower individuals to produce content for a global audience without restrictive barriers.
- Establishing common 3D formats and exchanges. The standardization of engines and communications suites is fairly complex in comparison to how 3D-objects conventions will emerge. Thus, it is essential to guarantee not only software interoperability, but also the compatibility of the formats used to create 3D content.

2) Decentralization

In the ever-evolving landscape of digital worlds, the challenge of decentralization is key, intertwining the realms of the Industrial Metaverse and Web3, which envisions a decentralized Internet orchestrated by independent developers and users, catalyzed by technologies like blockchain [262].

Though distinct, the Industrial Metaverse and Web3 may tread parallel evolutionary paths. The technological transitions ushered in by the Industrial Metaverse can align with the principles of Web3, fostering societal changes as the ones targeted by the Industry 5.0 paradigm, thus empowering individual consumers and emerging companies. Notably, several authors have already embraced blockchain as a key technology for the Industrial Metaverse, especially in relation to digital asset management [132].

The principles that define Web3, such as decentralization and a shift of online power to users, are poised to play a critical role in shaping a thriving both Commercial and Industrial Metaverses. Such a synergy not only promotes healthy competition but also ensures that the Metaverse construction mirrors the decentralized dynamics of the physical world, driven by independent users, developers and businesses.

Web3 introduces considerations of trust, asserting that centralized models mask the authenticity of virtual entitlements. By embracing decentralized databases and servers, Web3 proponents indicate that trust is inherently strengthened, laying a robust foundation for the Metaverse health and prospects [166].

In addition, decentralized computing can be useful for the Industrial Metaverse, as it was previously described in relation to opportunistic Edge Computing (some authors have already proposed to make use of Edge Computing devices to create distributed Metaverses [74]). By leveraging the computing power of devices deployed throughout industrial facilities, the paradigm shifts towards sharing processing capabilities. Specifically, the advent of blockchain technology emerges as a catalyst for decentralized computing, providing both the technological mechanisms and the economic model [263], [264], [265].

3) Legal challenges: data protection and GDPR

The advent of Commercial and Industrial Metaverses brings forth multiple legal challenges, particularly concerning data protection and compliance with regulations such as the General Data Protection Regulation (GDPR). As participants engage in virtual environments, the collection and processing of vast amounts of personal information will become central to the functioning of the Metaverse [266].

Participants in the Metaverse generate a wealth of personal data, including physiological reactions, movements and even brainwave patterns [267]. Although the continuous monitoring of user behavior allows Industrial Metaverse developers to tailor services with a high degree of precision, this constant data collection poses significant data security responsibilities for businesses operating within an Industrial Metaverse. Specifically, the immersive nature of the Industrial Metaverse enables companies to gather information about participants seamlessly. For example, observing a Meta-Operator consistently visiting vending machines may lead to the assumption that the operator is not efficient or that he/she can have potential health problems related to the kind of food he/she is consuming. The challenge here lies in determining how this information is collected, used and whether operators need to actively provide consent for such a data processing.

The GDPR imposes specific obligations on entities based on their role as either a 'controller' or a 'processor' of personal data. Determining these roles within the complex structure of the Industrial Metaverse becomes challenging. Questions arise regarding whether a single primary administrator governs all data processing or if multiple entities within the Industrial Metaverse have distinct data processing objectives.

Moreover, establishing a clear framework for data protection within the Industrial Metaverse involves unraveling a complex web of relationships. Key questions include whether there is a centralized authority for data collection and decision-making, or if multiple entities independently harvest personal data. This complexity raises concerns about privacy notices, user consent, liability in case of data breaches, and the formulation and enforcement of data-sharing agreements.

Furthermore, determining how various entities that interact in the Industrial Metaverse should display privacy notices to Meta-Operator and obtain their consent becomes a critical issue. The question is whether privacy notices should be presented collectively and how can Meta-Operator consent can be effectively obtained in an immersiveenvironment.

#### 4) Export and data localization

Exporting and localizing data in the Industrial Metaverse pose challenges related to the seamless movement of information across borders. The following are the main considerations to be taken into account in this regard:

- Seamless data movement. Achieving 'seamlessness' in the Industrial Metaverse requires the quick and frictionless movement of data across geographical and jurisdictional barriers. However, as data export and localization regulations become more stringent, ensuring seamless data transfer becomes challenging.
- Legal considerations. Legal frameworks, such as the European Judicial ruling, impact data exporters, requiring them to assess whether the destination country has legislation to adequately secure data in compliance with EU requirements. Data localization laws in various countries may impose restrictions on the movement of data across borders, posing legal challenges for Metaverse managers.
- Data localization legislation. Many countries implement 'data localization' legislation, which imposes specific restrictions on data leaving the country of origin. Such laws aim to ensure that data generated within a particular jurisdiction remain within their boundaries. Navigating these legal restrictions while enabling cross-border data flow in the Industrial Metaverse presents a significant problem.
- Data security measures. With varying regulations on data security and privacy, implementing standardized data security measures that satisfy the requirements of different jurisdictions becomes a complex task. Striking a balance between global interoperability and local compliance is crucial for the success of data export strategies.
- International cooperation. The Industrial Metaverse requires international cooperation and dialogue to address the legal complexities associated with data export and localization. Standardizing practices, fostering collaboration among stakeholders and engaging with regulatory bodies globally will be essential to create a framework that enables seamless data movement while respecting legal boundaries.

#### 5) Intellectual property

The intellectual property rights (IPR) in the Industrial Metaverse introduce unique challenges, including issues of ownership, joint authorship and the intersection of property rights.

First, it must be noted that determining ownership of intellectual property rights in collaborative Industrial Metaverses poses challenges. When multiple stakeholders contribute to the creation of intellectual property, issues of joint authorship and co-ownership arise. Defining the rights and responsibilities of each participant becomes complex in an interactive XR environment. Traditional risk assessment models may need enhancements to effectively evaluate potential conflicts and breaches arising from the combination of property rights.

In addition, the traditional scope of use terms in IPR licenses may need to be adapted to account for the diverse and dynamic nature of the Industrial Metaverse. Stakeholders must carefully define how intellectual property can be used, especially when it involves collaboration and the integration of different creative elements.

Finally, it must be noted that some authors have already analyzed the problem of copyright infringement in Metaverse environments, stating that such environments will not involve a paradigm shift in copyright law, so the developers and users of Metaverse platforms can keep on obeying existing copyright rules and copyright contract law [268].

#### 6) E-Commerce regulations

E-commerce regulations governing the Industrial Metaverse should consider their compliance with existing and evolving frameworks. For instance, the following regulations need to be considered in an European context:

- Platform to Business Regulation (P2B Regulation). The EU P2B Regulation [269], designed to address unfair trading practices in online intermediary services, is crucial for ensuring fairness and transparency in the Metaverse e-commerce components. Requirements include explaining differentiated treatment of goods, providing reasons for discontinuing a vendor's participation and disclosing criteria for ranking products and services in search results.
- Digital Services Act (DSA). The DSA regulation of the European Commission [270] aims to enhance consumer transparency and safety in online settings while supporting innovation in digital firms. Key provisions include defining 'illegal information' and 'illegal behavior', expanding the scope of covered online services, increasing intermediaries' liability, assigning responsibility for prompt removal of illegal/harmful information and promoting transparency in internet advertising and smart contracts. Striking a balance between holding digital intermediaries accountable and avoiding unjustified fines for service providers is a challenge.
- Digital Markets Act (DMA). The DMA [271] seeks to identify gatekeeper platforms and establish a new framework, requiring or prohibiting certain gatekeeper practices. DMA grants the European Commission investigative powers and enforcement capabilities for behavioral and structural solutions, including divestitures. Gatekeeper platforms with significant user reliance, lock-in effects, and data-driven advantages may face classification under the DMA. This has implications for key Industrial Metaverse players.

*7) EU AI regulation*

The European Commission has recently passed an AI Regulation to address the use of artificial intelligence [272], which is expected to play a significant role in facilitating human interactions within the Industrial Metaverse. The regulation aims to regulate specific AI techniques and imposes various duties on both suppliers and users of such AI systems. Specifically, the regulation identifies and prohibits certain AI techniques, particularly those deemed high-risk [273]. Suppliers and users of high-risk AI systems within the Metaverse will be required to comply with specific duties outlined in the regulation. This can include adopting certain procedures and safeguards to mitigate risks associated with advanced AI technologies. In addition, stakeholders must ensure transparency in the deployment and operation of AI systems, especially those that contextualize or manipulate human responses and simulate reality, such as through 'deep fakes'.

## VII. FUTURE RESEARCH LINES

Figure 19 summarizes the main research lines for the creation of the future Industrial Metaverse that have been previously mentioned throughout this article. Ten main research lines are contemplated:

- High-performance computing. This research line involves all the different technologies that allow for performing more computing operations per time unit and with a higher efficiency. Thus, it includes the different improvements in efficient CPU/GPU processing, fast retrieval and storage techniques, large virtual-world persistence strategies or the different techniques to generate and visualize 3D content.
- High-performance cybersecurity. It includes the multiple research lines that deal with challenges related to securing the future Industrial Metaverse, like ones associated with biometric security, Dark Pattern prevention, low-power cybersecurity and post-quantum cybersecurity, as well as the different strategies to prevent attacks to virtual worlds.
- AI. It is related to the progress of the different AI disciplines that impact the Industrial Metaverse (e.g., ML, DL, FL), including data processing techniques (e.g., Edge-AI, Natural Language Processing (NLP), voice synthesis) and computer-vision techniques.
- Ultra-Low Latency Networks (ULLNs). This critical research line includes the different communication technologies that currently can provide low latencies (of less than 10ms), like 5G, 6G, WiFi 7 and WiFi 8.

- Regulations. More effort is still necessary to regulate all the aspects involved in the creation of the Industrial Metaverse, which must implicate not only industrial corporations and their suppliers, but also the multiple local, regional, state and international organizations that at some point impact industrial processes. This research line also includes the necessary standardization initiatives, which need to somehow regulate the communications, data exchanges and the overall interoperability of the future Industrial Metaverse.
- Meta-IoT/IIoT devices. This research area includes the different factors that allow for building Industrial Metaverse-ready IoT/IIoT devices, like the creation of energy-efficient devices, plug-and-play protocols, integration mechanisms or the development of smart edge-based processing techniques that lower IoT/IIoT computational load.
- Decentralized and distributed systems. More research is necessary to build a truly decentralized Industrial Metaverse with the help of Distributed Ledger Technologies (DLTs) like blockchain. In addition, topics of interest are the development of distributed solutions like Meta-operating systems or advanced ICPs and digitaltwins.
- OEC. The four main foundations of OEC systems need to be adapted and optimized to the Industrial Metaverse in order to facilitate the integration of IoT/IIoT devices and XR device intercommunications. As it was previously mentioned, such four main foundations involve peer discovery/routing, data routing and efficient resource sharing.
- New XR technologies. XR devices still need to evolve to improve current UX for industrial environments, so they need to become more comfortable and more energy efficient, provide better resolution/refresh rates/FoV and be more powerful. In addition, more effort has to be dedicated to developing very-light XR devices (e.g., smart contact lenses), to creating Shared Reality devices (with inherent non-cloud-dependent collaborative features) and to reduce the price of the devices in order to carry out massive industrial deployments.
- Advanced XR Accessories. Although XR accessories like gloves have evolved notably in the last years, other types of wearables are still in their infancy regarding their integration with the Industrial Metaverse, like EEG, EMG or BCI control devices. Moreover, such devices have to become energy efficient and be able to work jointly with other novel XR accessories, like advanced motion/gesture capture devices, haptic devices or other accessories that provide feedback by stimulating human senses.

## VIII. CONCLUSIONS

This article described the foundations of the Meta-Operator concept and provided useful guidelines for the future Industry 5.0 developers that will bring it to life. For such a purpose, this article described thoroughly the main components that will be required to forge future Meta-Operators, including the necessary XR devices and accessories, the development of opportunistic communication protocols and the integration with surrounding IoT/IIoT devices. In addition, this paper studied the essential parts of the Industrial Metaverse, the latest standardization initiatives and the different alternatives to deploy advanced architectures that will allow for providing immersive collaborative experiences. Furthermore, this article provided an extensive analysis on the main development, efficiency and legal challenges that future Industrial Metaverse developers will have to face in the years to come. Thus, this paper provided a holistic view on three concepts (Industrial Metaverse, Meta-Operators, Industry 5.0) that together will pave the way for the creation of the next generation of smart factories.

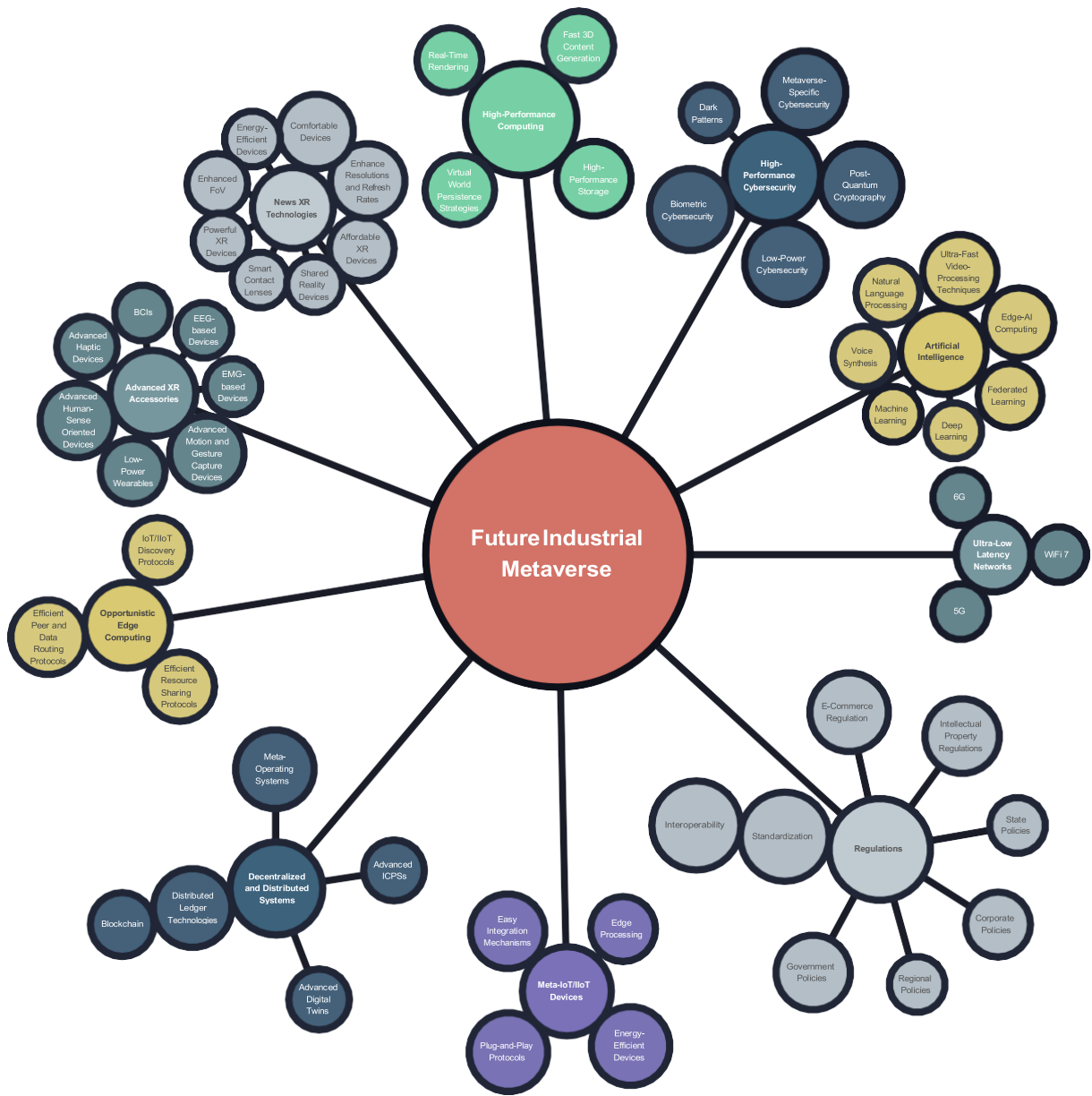

FIGURE 19: Summary of the most promising research lines for creating the Industrial Metaverse.

Tiago M. Fernández-Caramés, Paula Fraga-Lamas: Forging the Industrial Metaverse

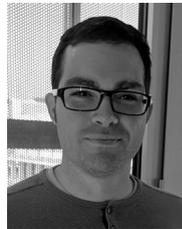

TIAGO M. FERNÁNDEZ-CARAMÉS (S'08-M'12-SM'15) works as an Associate Professor at the University of A Coruña (UDC) (Spain), where he obtained his MSc degree and PhD degrees in Computer Science in 2005 and 2011. Since 2005 he has worked in the Department of Computer Engineering at UDC: from 2005 to 2009 through different predoctoral scholarships and, in parallel, since 2007, as a professor. His current research interests include XR technologies, IoT/IIoT systems and blockchain, as well as the other different technologies involved in the Industry 4.0 and 5.0 paradigms. In such fields, he has contributed to more than 120 papers for conferences, high-impact journals and books, as well as to multiple patents.

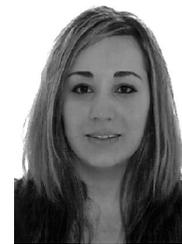

PAULA FRAGA-LAMAS (M'17-SM'20) received the M.Sc. degree in Computer Engineering from the University of A Coruña (UDC), in 2009, and the M.Sc. and Ph.D. degrees in the joint Mobile Network Information and Communication Technologies Program from five Spanish universities, University of the Basque Country, University of Cantabria, University of Zaragoza, University of Oviedo, and University of A Coruña, in 2011 and 2017, respectively. She holds an MBA and postgraduate studies in business innovation management (Jean Monnet Chair in European Industrial Economics, UDC), and Corporate Social Responsibility (CSR) and social innovation (Inditex-UDC Chair of Sustainability). Since 2009, she has been with the Group of Electronic Technology and Communications (GTEC), Department of Computer Engineering (UDC). She has over 100 contributions in indexed international journals, conferences, and book chapters, and holds 3 patents. She has also been participating in over 40 research projects funded by the regional and national government as well as R&D contracts with private companies. She is actively involved in many professional and editorial activities, acting as reviewer of more than 35 international journals, advisory board member, topic/guest editor of top-ranked journals and TPC member of international conferences. Her current research interests include mission-critical scenarios, Industry 5.0, Internet of Things (IoT), Cyber-Physical Systems (CPS), Augmented Reality (AR), fog and edge computing, blockchain and Distributed Ledger Technology (DLT), and cybersecurity.